\documentclass{aa}

\usepackage{graphicx}
\usepackage{txfonts}
\usepackage[colorlinks=true,linkcolor=blue]{hyperref}
\hypersetup{citecolor=blue, urlcolor=blue}

\usepackage{amsmath}    
\usepackage{subfig}
\usepackage[normalem]{ulem}
\usepackage[separate-uncertainty=true]{siunitx}
\usepackage{dcolumn}
\usepackage{arydshln}    
\usepackage{tablefootnote}
\usepackage[dvipsnames]{xcolor}
\usepackage{orcidlink}
\usepackage{makecell}
\usepackage[flushleft]{threeparttable}
\usepackage{soul}
\usepackage{pdflscape}
\usepackage{float,lscape}

\newcommand{\g}{G310.7--5.4}
\newcommand{\snr}{Abeona}

\newcommand{\ps}{\texttt{PointSource}}

\newcommand{\lp}{\texttt{LogParabola}}

\newcommand{\pss}{\texttt{PS}}
\newcommand{\pls}{\texttt{PL}}
\newcommand{\lps}{\texttt{LP}}

\usepackage{scalerel}

\newcommand\hii{H\,\protect\scaleto{$II$}{1.2ex}}

\newcommand{\fermi}{\textit{Fermi}}
\newcommand{\fermilat}{\textit{Fermi}--LAT}

\DeclareSIUnit\erg{erg}
\DeclareSIUnit\SIsigma{\ensuremath{\sigma}}
\DeclareSIUnit\parsec{pc}
\DeclareSIUnit\photon{ph}
\DeclareSIUnit\jansky{Jy}
\DeclareSIUnit\beam{beam}
\DeclareSIUnit\sterrad{sr}
\DeclareSIUnit\year{yr}
\DeclareSIUnit\radians{rad}

\newcommand{\noop}[1]{}

\begin{document}

   \title{Radio detection of supernova remnant \g\ \\ with $\gamma$-ray counterpart: \snr\ SNR}

   \subtitle{}

   \author{
          Christopher Burger-Scheidlin\,\inst{1,2,3}\thanks{Email: \href{mailto:cburger@cp.dias.ie}{cburger@cp.dias.ie}}\orcidlink{0000-0002-7239-2248}
          \and
          Brianna D. Ball\,\inst{4}\orcidlink{0009-0003-2088-9433}
          \and
          Sanja Lazarevi\'c\,\inst{2,5,6}\orcidlink{0000-0001-6109-8548}
          \and 
          Roland Kothes\,\inst{7,4}\orcidlink{0000-0001-5953-0100}
          \and \\
          Robert Brose\,\inst{8}\orcidlink{0000-0002-8312-6930}
          \and
          Jonathan Mackey\,\inst{1,3}\orcidlink{0000-0002-5449-6131}
          \and
          Miroslav D. Filipovi\'c\,\inst{2}\orcidlink{0000-0002-4990-9288}
          \and
          Zachary J. Smeaton\,\inst{2}\orcidlink{0009-0009-7061-0553}
          \and \\
          Andrew M. Hopkins\,\inst{9}\orcidlink{0000-0002-6097-2747}
          \and
          Denis Leahy\,\inst{10}\orcidlink{0000-0002-4814-958X}
          \and
          Mehrnoosh Tahani\,\inst{11,12}\orcidlink{0000-0001-8749-1436}
          \and
          Jennifer L. West\,\inst{7,13}\orcidlink{0000-0001-7722-8458}
          \and
          Tayyaba Zafar\,\inst{9}\orcidlink{0000-0003-3935-7018}
          }
    
   \institute{
        Astronomy \& Astrophysics Section, School of Cosmic Physics, Dublin Institute for Advanced Studies, DIAS Dunsink Observatory, Dublin D15 XR2R, Ireland
        \and
        Western Sydney University, Locked Bag 1797, Penrith South DC, NSW 2751, Australia
        \and
        School of Physics, University College Dublin, Belfield, Dublin D04 V1W8, Ireland
        \and
        Department of Physics, University of Alberta, 4-181 CCIS, Edmonton, Alberta T6G 2EI, Canada
        \and
        ATNF, CSIRO, Space and Astronomy, PO Box 76, Epping, NSW 1710, Australia
        \and
        Astronomical Observatory, Volgina 7, 11060 Belgrade, Serbia
        \and
        Dominion Radio Astrophysical Observatory, Herzberg Astronomy \& Astrophysics, National Research Council Canada, P.O. Box 248, Penticton, BC V2A 6J9, Canada
        \and
        Institute of Physics and Astronomy, University of Potsdam, 14476 Potsdam-Golm, Germany
        \and
        School of Mathematical and Physical Sciences, 12 Wally's Walk, Macquarie University, NSW 2109, Australia
        \and
        Department of Physics and Astronomy, University of Calgary, Calgary, Alberta, Canada
        \and
        Department of Physics \& Astronomy, University of South Carolina, Columbia, SC 29208, USA
        \and
        Kavli Institute for Particle Astrophysics \& Cosmology (KIPAC), Stanford University, Stanford, CA 94305, USA
        \and
        School of Natural Sciences, University of Tasmania, PO Box 807, Sandy Bay, TAS 7006, Australia
        }

    \titlerunning{Detection of \g: \snr\ SNR}
    \authorrunning{C. Burger-Scheidlin et al.}

  \abstract
   {\g\ is a supernova remnant (SNR) candidate identified as a faint shell in the second epoch \textit{Molonglo Galactic Plane Survey} (MGPS-2), but this has not been followed up with multi-wavelength observations until now. It is an example of an SNR at high Galactic latitude showing spatially coinciding $\gamma$-ray emission.}
   {Here, we make the first detailed investigation of the radio emission from the \g\ region, aiming to characterise the radio structure, polarisation measurements and the coinciding GeV emission.}
   {We used recent radio continuum observations at \SI{943.5}{\mega\hertz} from the \textit{Evolutionary Map of the Universe} (EMU) and the \textit{Polarisation Sky Survey of the Universe's Magnetism} (POSSUM) surveys with the \textit{Australian Square Kilometre Array Pathfinder} (ASKAP), as well as 16.5 years of \fermilat\ observations. We furthermore considered the multiwavelength context of the object by investigating observations previously conducted with other instruments, such as infrared and X-ray surveys.}
   {We confirm the SNR candidate as a new supernova remnant, dubbed \snr. We detect the presence of a faint, extended, bilateral radio shell of the size of around \SI{30}{\arcmin} diameter and ASKAP radio flux density of $1.5^{+1.5}_{-0.1}$\,\SI{}{\jansky} with no obvious infrared counterparts. With a radio surface brightness of about $2.4^{+2.4}_{-0.1}\times10^{-22}\,\SI{}{\watt\per\square\meter\per\hertz\per\sterrad}$, this SNR is one of the faintest radio SNRs known. The northern part of the shell shows linearly polarised radio emission, characteristic of synchrotron emission in SNRs. The physical size of the SNR is estimated to be around $42^{+42}_{-21}\,\SI{}{pc}$, which would give a distance of around $4.9^{+4.9}_{-2.5}\,\SI{}{\kilo pc}$. Furthermore, the spatially coincident $\gamma$-ray source 4FGL\,J1413.9--6705 shows an energy flux of \SI{1.26\pm0.35e-6}{\mega\electronvolt\per\cm\squared\per\s} with a significance of \SI{5.7}{\SIsigma} between \SI{100}{\mega\electronvolt} and \SI{100}{\giga\electronvolt}.
   The SNR is also put in context with known high-latitude SNRs with $\gamma$-ray counterparts and compared with their observational properties.
   }
   {New radio surveys are discovering a population of faint high-Galactic-latitude SNRs with GeV counterparts in isolated environments.  More sensitive observations with the \textit{Cherenkov Telescope Array Observatory} (CTAO) in the next few years should provide valuable insights into particle acceleration and escape in these important cosmic ray accelerators.}


   \maketitle
%

\section{Introduction}

\begin{figure*}[h!]
    \centering
    \captionsetup[subfigure]{labelformat=empty}
    \subfloat[\centering]{{\includegraphics[width=0.915\linewidth, trim={0.85cm 1.1cm 1.93cm 2.5cm}, clip]{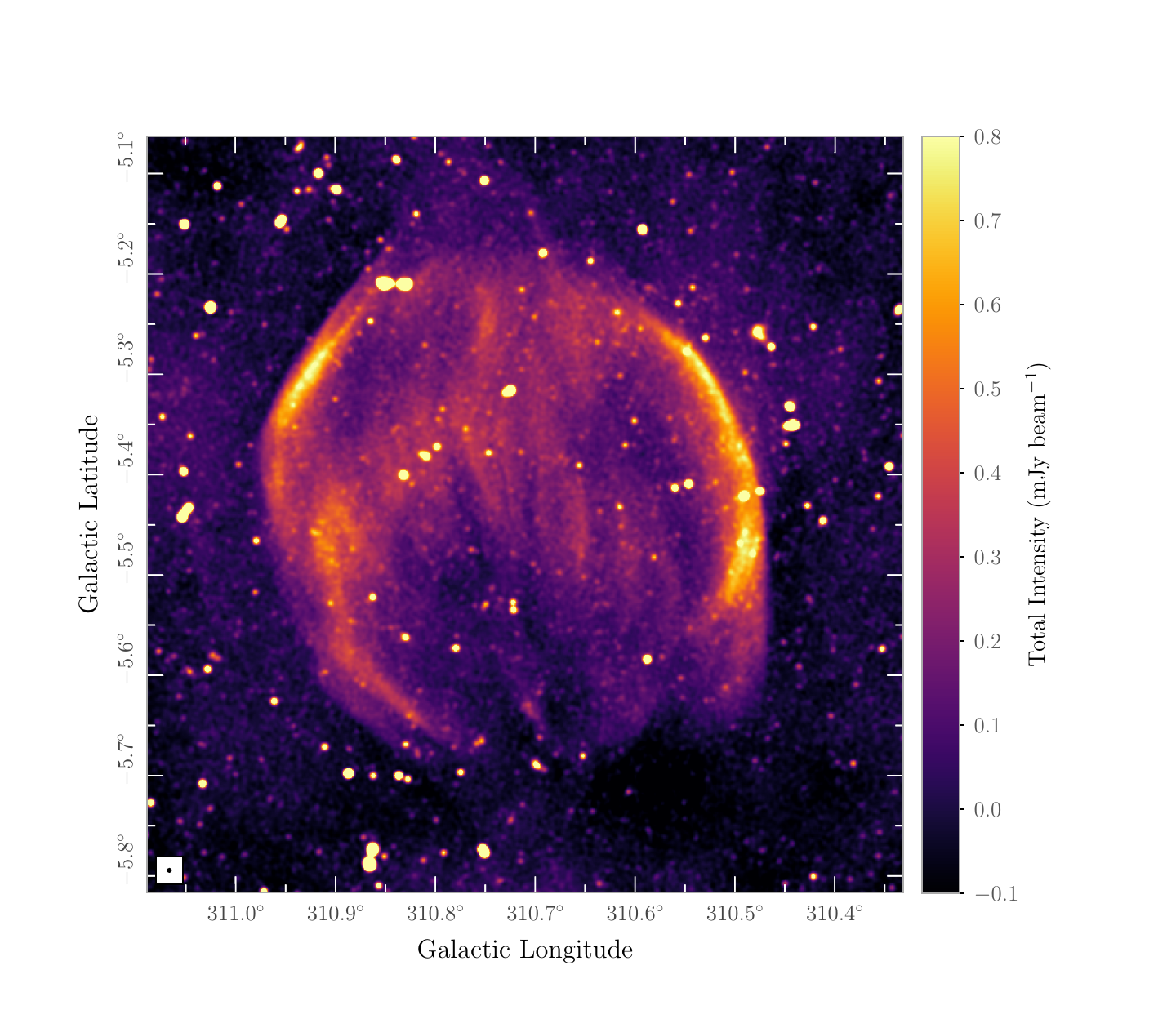}}}%
    \vspace{-0.4cm}
    \caption{ASKAP 943.5 MHz total intensity image of Abeona (\g). The synthesised beam of the image is $\SI{15}{\arcsec}$, represented with a small black circle in the bottom-left corner. The emission is centred around $l, b$ = 310.73\SI{}{\degree}\,,\,--5.43\SI{}{\degree} (R.A./Dec =14:12:34\,/\,--67:04:03) with a diameter of approximately \SI{30}{\arcmin} ($\sim\SI{15}{\arcmin}$ radius) and low eccentricity.}
    \label{fig:askap-i}
\end{figure*}

Studying supernova remnants (SNRs) is essential when attempting to understand the enrichment of the interstellar medium (ISM), the evolution of the Milky Way, the origins of Galactic cosmic rays (CRs), and the dynamics of the turbulent ISM.
The population of SNRs is known to be substantially under-represented~\cite[i.e.][]{Brogan_2006, GreenA_2014, Ranasinghe_2023ApJS..265...53R, Ball_2023} and to date, around 310 SNRs are confirmed~\citep{Ferrand_2012, Green_2017, Hurley-Walker_2019PASA...36...45H, Green_2025JApA...46...14G}, the majority of which are first detected with radio surveys. Nearly half of those are later detected in X-rays, as well as one third in optical~\citep{Green_2025JApA...46...14G}. Around \SI{10}{\percent} are later detected at high energies (HE) in $\gamma$-rays~\citep{Acero_2016ApJS..224....8A}, and even fewer at very high energies~\citep[e.g.][]{HESS_2018A&A...612A...3H}.

The \textit{Evolutionary Map of the Universe}~\cite[EMU;][]{EMU_2022, Hopkins_2025PASA...42...71H} survey from the \textit{Australian Square Kilometre Array Pathfinder}~\citep[ASKAP;][]{Hotan_2021} aims to map the entire southern sky with high resolution at the central radio frequency of \SI{943.5}{\MHz}. With ongoing observations, more and more extended faint Galactic objects such as the SNR presented in this work are being revealed~\citep[e.g.][]{Dubner_2015A&ARv..23....3D, Smeaton_2024MNRAS.534.2918S, 2025arXiv250504041F, Ball_2025ApJ...988...75B}.

The origin of CRs has been in discussion since their detection by~\citet{Hess_1912}. It is commonly accepted that CRs with energies up to around \SI{3e15}{\electronvolt} are of Galactic origin~\citep[i.e.][]{Aloisio_2012APh....39..129A, Globus_2015PhRvD..92b1302G}. SNRs were first suggested to be CR accelerators by \citet{Baade_and_Zwicky_1934} and have been found to be substantially contributing to the Galactic cosmic ray spectrum, especially at GeV -- TeV energies~\cite[i.e.][]{Koyama_1995, Ackermann_2013}.

Non-thermal synchrotron emission is the main process for radio emission in SNRs~\citep[e.g.][]{Berezhko_2004,book2,book1} produced by highly relativistic particles accelerated at the shock front caused by the expanding shell, as postulated by diffusive shock acceleration theory~\citep[e.g.][]{Blandford_1978, Bell_1978}.

To classify new SNRs as such and distinguish them from other, morphologically similar objects such as \hii\ regions, it is beneficial to show a lack of spatial correlation with strong (mid-) infrared (IR) emission~\citep{Whiteoak_1996,Cotton_2024MNRAS.529.2443C,Anderson_2025A&A...693A.247A}. To then establish a firm detection, evidence of polarised radio emission from the underlying synchrotron process, measurements of the spectral index in agreement with expectations for synchrotron emission, or multiwavelength detection, such as conclusive evidence in optical or X-rays are required~\citep[e.g.][]{Green_2025JApA...46...14G, Ball_2025ApJ...988...75B}.

In recent years, SNRs at high Galactic latitudes have received some attention, as more objects are detected off the Galactic plane at radio wavelengths~\citep[i.e.][]{Kothes_2017A&A...597A.116K, Filipovic_2022MNRAS.512..265F, Jing_2025ApJ...980..162J, Lazarevic_in_prep}. Amongst others, G150+4.5~\citep{Devin_2020}, G17.8+16.7~\citep{Araya_2022}, Calvera SNR~\citep[G118.4+37.0;][]{Arias_2022, Araya_2023}, and Ancora SNR~\citep[G288.8--6.3;][]{Filipovic_2023, Burger-Scheidlin_2024} were detected at high latitudes in $\gamma$-rays as well. These SNRs expand in low-density environments with low background noise and diminished risk of source confusion for $\gamma$-ray observations. The environment also makes it easier to detect faint diffuse emission in radio and other wavelengths, like X-rays.
Recent discoveries at optical wavelengths, together with counterparts in $\gamma$-rays such as Nereides~\citep[G107.7‑5.1;] []{Fesen_2024, Araya_2024} demonstrate that optical detections of SNRs are possible, despite challenges with very faint optical signals due to extinction effects.

\g\ was first observed and suggested as an SNR candidate by~\citet{GreenA_2014}. Recently, it was also classified as an SNR candidate by the Murchison Widefield Array~\citep[MWA;][]{Mantovanini_2025PASA...42...21M}. Here, we report the confirmation of the supernova remnant, dubbed \snr\footnote{Abeona, the goddess of outward journeys in Roman mythology, protected travellers on their departing paths. This SNR, its progenitor having wandered off the Galactic plane and into the Galactic halo, therefore carries her name.}, at high Galactic latitude, using EMU/POSSUM survey data from ASKAP.
We analysed all relevant available data of the region, including radio intensity, polarisation, and gamma-ray emission and elaborated on data processing for this study (Sect.~\ref{sec:observations}). We present our results in Sect.~\ref{sec:results}. In Sect.~\ref{sec:discussion}, we discuss the implications of our results, give a distance estimate, and compare \g\ with other SNRs at high Galactic latitudes. Finally, we give our conclusions in Sect.~\ref{sec:conclusion}.

\section{Observations and data processing}
\label{sec:observations}

\subsection{Australian Square Kilometre Array Pathfinder (ASKAP)}

In this study, we make use of ASKAP observations~\citep{Hotan_2021} from two projects: \textit{Evolutionary Map of the Universe}~\citep[EMU\footnote{\href{http://hdl.handle.net/102.100.100/479788?index=1}{ASKAP Data Products for Project AS201 (EMU)}};][]{Norris_2011, Norris_2021, EMU_2022, Hopkins_2025PASA...42...71H} and \textit{Polarisation Sky Survey of the Universe’s Magnetism}~\citep[POSSUM\footnote{\href{http://hdl.handle.net/102.100.100/479787?index=1}{ASKAP Data Products for Project AS203 (POSSUM)}};][]{Gaensler_2010, Gaensler_2025PASA...42...91G}. ASKAP, a radio interferometer with baselines of up to \SI{6}{\km}, consists of 36 twelve-metre dishes and is located in Western Australia. Each antenna forms 36 simultaneous beams by utilising a phased array feed, which provides ASKAP with a very wide field of view of approximately \SI{30}{deg \squared}. This enables large-scale sky observations, making the array well-suited for use in surveys. 

Stokes~$I$ images are taken from the EMU survey and were observed at a central frequency of \SI{943.5}{\MHz}, with a bandwidth of \SI{288}{\MHz} divided into 288 individual channels of \SI{1}{\MHz} width each. Images are then convolved to a common angular resolution of \SI{15}{\arcsec} with a median root-mean-square noise of \SI{30}{\micro\jansky\per\beam}~\citep{Hopkins_2025PASA...42...71H}. The standard \texttt{ASKAPsoft}~\citep{Guzman_2019} pipeline is used for processing observations, following~\citet{Norris_2021}.

Stokes $Q$ and $U$ data cubes are taken from the POSSUM survey. They have identical observing characteristics and are observed simultaneously with Stokes~$I$. All channels in the Q and U data cubes are convolved to a common resolution of \SI{18}{\arcsec}. To obtain intrinsic Stokes~$Q$, $U$, polarised intensity (PI) and polarisation angle (PA) images, we utilised the rotation measure (RM) synthesis technique. Instead of the more common Fourier transform method, we determine the Faraday depth (FD) function for each pixel by de-rotating the $Q$ and $U$ data at each frequency for rotation measures in the range of $-2000$ to $+2000$~rad\,m$^{-2}$. This technique is described in more detail in \citet{Ball_2023}. Resulting maps for Stokes~$Q$, $U$, PI, PA, and RM are taken from the peak of the FD function in each pixel.

In total, four fields from the EMU/POSSUM survey were used: \texttt{EMU\,1442-64} (\texttt{SB\,45821}), \texttt{EMU\,1356-64} (\texttt{SB\,53310}), \texttt{EMU\,1418-69} (\texttt{SB\,62565}), and \texttt{EMU\,1413-70} (\texttt{SB69812}). Here, SB refers to the scheduling block of the observation. These fields were observed between Nov 24, 2022 and Dec 25, 2024. To utilise all available ASKAP pipeline data and enhance the signal-to-noise ratio, these four fields were mosaicked together using the corresponding weight files and cubes.

All pipeline-made, science-ready data are available on the CSIRO ASKAP Data Science Archive (CASDA)\footnote{\href{https://research.csiro.au/casda/}{research.csiro.au/casda/}}.

\begin{figure}
    \captionsetup[subfigure]{labelformat=empty}
    \subfloat[\centering]{{\includegraphics[width=1\columnwidth, trim={0.5cm 1.4cm 0.93cm 1.6cm}, clip]{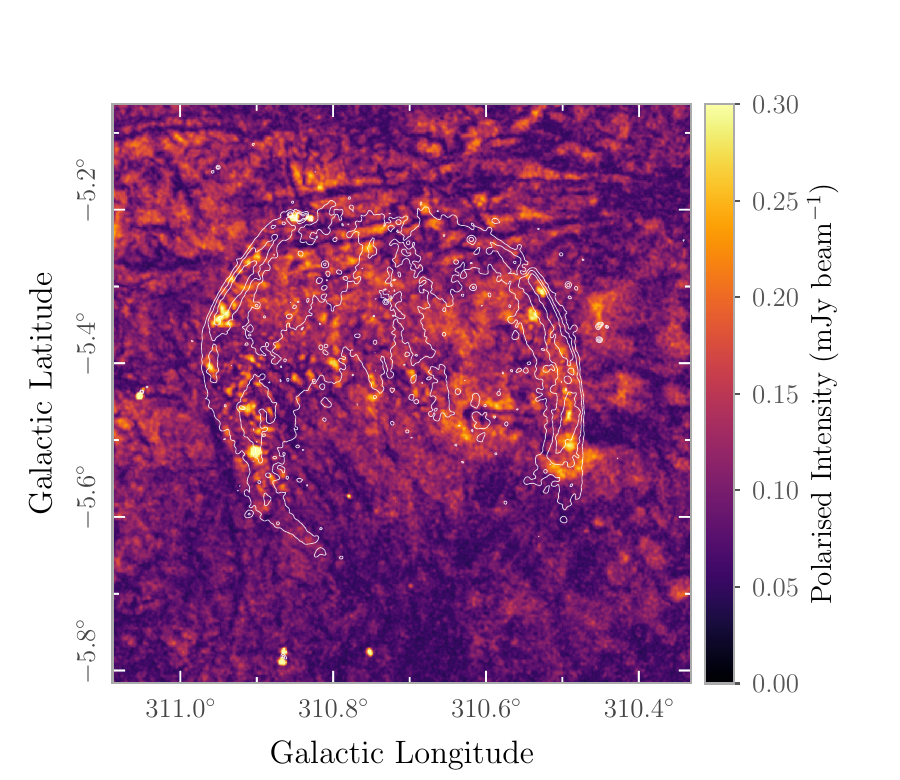}}}%
    \vspace{-0.6cm}
    \par
    
    \subfloat[\centering]{{\includegraphics[width=1\columnwidth, trim={0.5cm 1.4cm 0.93cm 1.6cm}, clip]{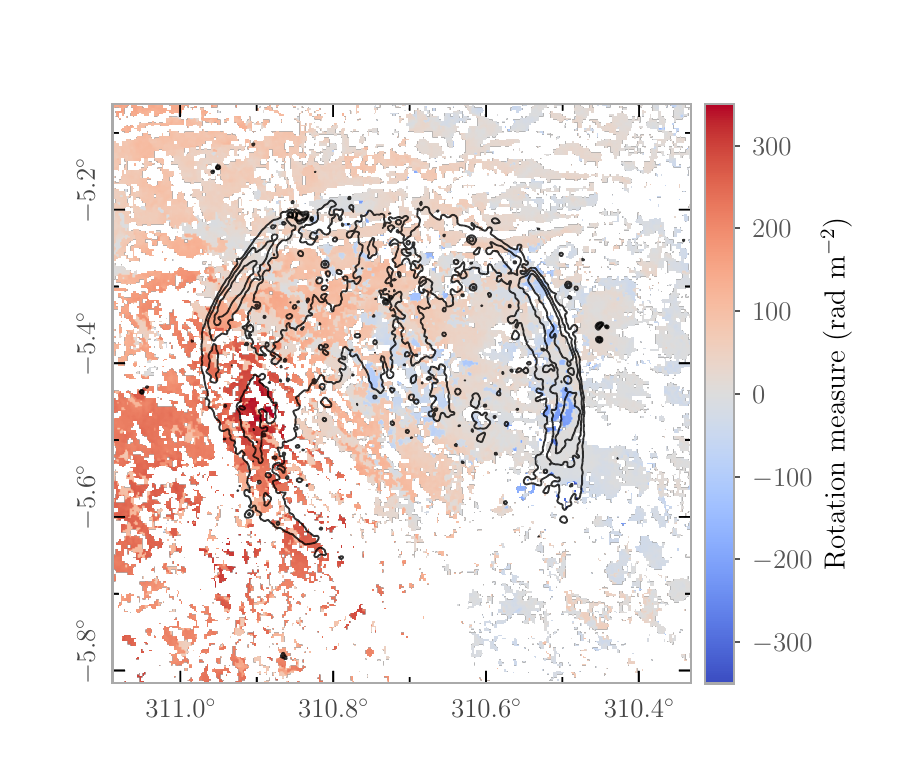}}}%
    \vspace{-0.6cm}
    \par
    
    \subfloat[\centering]{{\includegraphics[width=1\columnwidth, trim={0.5cm 0.1cm 0.93cm 1.6cm}, clip]{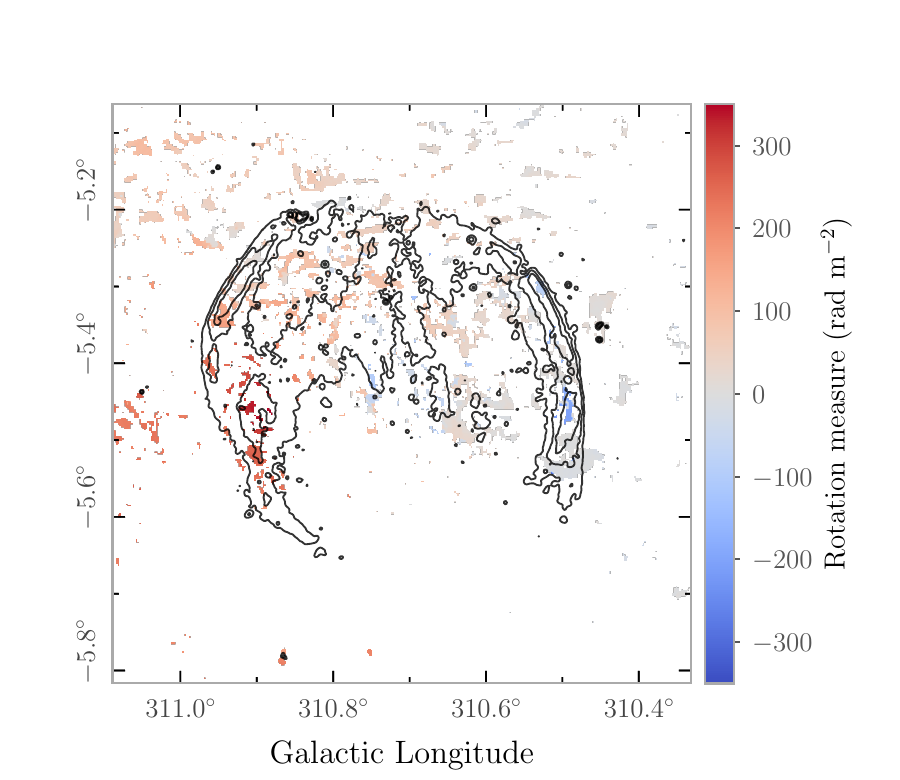}}}%
    \vspace{-0.4cm}
    
    \caption{\textbf{Top:} Polarised intensity image, overlaid with total intensity contours (white). \textbf{Centre:} Rotation measure, overlaid with total intensity contours (black) in the range of $-350$ to $+\SI{350}{\radians\per\square\meter}$, masking pixels with significances below a \SI{6}{\SIsigma} threshold. \textbf{Bottom:} Same as the middle, except with a \SI{10}{\SIsigma} threshold cut instead of \SI{6}{\SIsigma}. 
    }
    \label{fig:askap-pi}
\end{figure}

\subsection{\fermilat}

For the $\gamma$-ray analysis, data from the \textit{Large Area Telescope} (LAT; \citealt{Atwood_2009ApJ...697.1071A, Ackermann_2012ApJS..203....4A}) aboard the \textit{Fermi} observatory, a space-based $\gamma$-ray telescope launched in 2008, was used. In total, \SI{16.5}{\year} of \fermilat\ Pass 8 data from August 2008 to February 2025\footnote{\num{239557417}\:--\:\num{761514810} \fermi\ mission elapsed time (MET)} were analysed. A radius of \SI{20}{\degree} in the region of interest (ROI) centred on the radio SNR at Galactic position $l,~b$\footnote{R.A./Dec = 14:12:23.8\,/\,--67:04:13.3}~$ =\SI{310.7}{\degree}$, $\SI{-5.4}{\degree}$ was selected in the energy range of 100~MeV to 1~TeV. We used the latest version of \textit{fermipy} (version 1.2.2; \citealt{Fermipy_2017}) and \textit{fermitools} v2.2\footnote{\href{https://fermi.gsfc.nasa.gov/ssc/data/analysis/documentation/}{fermi.gsfc.nasa.gov/ssc/data/analysis/documentation/}} software packages to analyse $\gamma$-ray data, together with the latest instrument response function $\texttt{P8R3\_SOURCE\_V3}$~\citep{Atwood_2013}.

To compensate for poor angular resolution at low energies, we used a joint, four-component approach, adapted from~\citet{Abdollahi_2020ApJS..247...33A}, where different point-spread-function (PSF) classes (0-3) are modelled as independent sub-selections and then joined to one overall model that optimises on the log-likelihood of the individual models. Recommended filters ($\texttt{DATA\_QUAL}>0 \; \texttt{\&\&} \; \texttt{LAT\_CONFIG}==1$) were applied for data reduction, as well as standard zenith angle cuts of \SI{90}{\degree} to mitigate atmosphere-related effects.

A pixel size of \SI{0.1}{\degree} with eight logarithmic energy bins per decade was used. The underlying source catalogue was the fourth \fermi\ catalogue~\citep{Abdollahi_2020ApJS..247...33A}, specifically the incremental Data Release~4 version 4FGL--DR4 ~\cite[v35;][]{Ballet_2024_Fermi-4FGL-DR4}, modelling sources that were lying up to \SI{5}{\degree} outside the ROI. The Galactic diffuse emission model \texttt{gll\_iem\_v07.fits} with an isotropic component corresponding to the chosen event class (\texttt{iso\_P8R3\_SOURCE\_V3\_v1.txt}) was applied to account for diffuse emission.

The normalisation parameters of sources within \SI{3}{\degree}, and spectral parameters within \SI{2}{\degree}, were left to vary during fitting, as were the normalisation factors of sources above a test statistic,\footnote{TS is defined as twice the difference between log-likelihoods of source plus background ($\mathcal{L}_1$) and only background as a null hypothesis ($\mathcal{L}_0$):  $\text{TS} =2\,\text{ln}(\mathcal{L}_1/\mathcal{L}_0$).} $\text{TS}=10$ within the ROI. All other source parameters were kept fixed to their 4FGL--DR4 catalogue values. Furthermore, an energy dispersion correction was applied to all sources, except for the isotropic diffuse emission background model.

We conducted three initial optimisations of the sources using the \texttt{optimize()} method, which provides a preliminary fit to bring parameters close to their maximum likelihood values. This was followed by an initial fit using the \texttt{fit()} method with the \texttt{newminuit} optimizer. Subsequently, sources with $\text{TS}<1$ were removed. Additional fitting revealed no new sources with $\text{TS}<1$.

As the 4FGL--DR4 catalogue shows a source spatially coinciding with \g, we focused on analysing the catalogue source, 4FGL~J1413.9--6705, subsequently referred to as J1413. Due to large-scale positive and negative residual emission at Galactic latitudes $\leq \SI{3}{\degree}$, possibly due to the Galactic diffuse model, an offset of \SI{1.5}{\degree} away from the Galactic plane was chosen for the centre maps position. Within \SI{1.0}{\degree} of \g, three separate 4FGL--DR4 catalogue sources can be found: 4FGL~J1413.9--6705 (\SI{0.18}{\degree} angular separation, overlapping with the south-eastern part of the radio shell), 4FGL~J1412.1--6631 (\SI{0.52}{\degree}) and  4FGL~J1420.7--6701 (\SI{0.84}{\degree}). All other sources are more than \SI{1.4}{\degree} from the centre of \g. Due to the relatively small extension of \snr\ compared to the PSF of \fermilat, paired with a very soft spectrum, no attempts were made to model the source with an extended spatial model. However, the \texttt{localize()} method was applied at energies above  \SI{1}{\GeV} to optimise the source position of J1413.

\subsection{Other energy bands}

A search through the archives of X-ray observatories such as XMM-Newton and Chanda returned no pointed observations that covered this field. No excess was seen in publicly available archival \textit{ROSAT} All Sky Survey Data X-ray observations~\citep[RASS;][]{Voges_1999}. No excess was seen at different X-ray energies in the publicly available \textit{eROSITA}-DE Data Release 1~\citep[DR1;][]{Sunyaev_2021}.
Thermal dust maps from Planck do not show any significant, structured emission overlapping with \g~\citep{Planck_2016}.

\textit{Wide-field Infrared Survey Explorer} \citep[WISE;][]{Wright_2010} data presented in this work were taken from the All-Sky Survey\footnote{\href{https://wise2.ipac.caltech.edu/docs/release/allsky/}{wise2.ipac.caltech.edu/docs/release/allsky/}}.

\section{Results}
\label{sec:results}

\subsection{Radio-continuum properties of \g}
\label{sec:results-radio-continuum}

A total intensity map from ASKAP at \SI{943.5}{\MHz} of the SNR is presented in Fig.~\ref{fig:askap-i}. We observe an extended, bilateral, low-surface brightness object, with thin shell-like emissions seen brightest in the western (right) part, as well as the north-eastern (top left) part of the object\footnote{All figures throughout this publication are presented in Galactic coordinates. Directions (north, east, south, west) are given with respect to that coordinate system.}.
The emission seems to be almost spherical, with some filaments towards the top and bottom right diverging from the circular shape. A second, smaller, and fainter ring of emission can be seen, shifted slightly to the south but almost fully contained within the larger shell. Additionally, broad and very dim filaments are visible in the central region. They partly follow the shape of the brighter shell-like features and are likely associated with the remnant.

The bilateral shape of SNRs is believed to be related to the structure of the ambient magneto-ionic medium into which the SNR is expanding. \citet{West_2016A&A...587A.148W} found a close relationship between the SNR's bilateral symmetry axis and the large-scale Galactic magnetic field. \citet{Cotton_2024ApJS..270...21C} published a catalogue of 36 high-latitude Galactic SNRs and found that most bilateral SNRs in their sample show a small angle between their symmetry axis and the Galactic plane of our Galaxy. As the large-scale magnetic field in our Galaxy is believed to be parallel to the plane at low latitudes, \citet{Cotton_2024ApJS..270...21C} identified those as Galactic plane SNRs. On the other hand, most of the other SNRs have an angle between their symmetry axis and the Galactic plane of close to \SI{90}{\degree}, indicating that those are halo SNRs. \g\ belongs to the latter group and therefore could be located in the Galactic halo.

The full diameter of \g\ is around \SI{30}{\arcmin}. Due to the low surface brightness, faint point-like sources, as well as a varying background, it is quite challenging to measure a reliable flux density. To resolve these challenges, we used methods described in~\citet{Ball_2023}, utilising the \textit{polygon\_flux} software developed by~\citet{Hurley-Walker_2019PASA...36...48H} for the GLEAM survey to handle extended sources with complicated backgrounds. The software allows for manual source region masking as well as user-defined subtraction of point sources and performs background subtraction. Averaging over eight individual measurements, and taking into account three different definitions of the background within the field of view for each measurement, resulted in 24 flux density calculations, which were then used to obtain the median value. We determined a flux density of $1.5^{+1.5}_{-0.1}\,\SI{}{\jansky}$ for \snr\ SNR. Uncertainties take into account statistical origins, as well as some systematic effects such as user-based region selection. Measurements of the bright top left region of the object show a peak diffuse intensity of around \SI{0.85}{\jansky\per\beam}, whereas the eastern region indicates a maximum around \SI{0.75}{\milli\jansky\per\beam}. A spectral index could not be determined due to the faint emission and instrument limitations.

\subsubsection{Polarised emission}

Polarised intensity (PI) maps show polarised emission throughout different regions of the shell up to \SI{0.3}{\milli\jansky\per\beam}, as presented in Fig.~\ref{fig:askap-pi} (top), such as the brightest regions of the north-eastern shell (around $l, b =$ \SI{310.9}{\degree}, \SI{-5.3}{\degree}) and the south-west tip of the emission (around $l, b =\SI{310.5}{\degree}, \SI{-5.5}{\degree}$). Using the same methods described in Sect.~\ref{sec:results-radio-continuum}, we obtain a polarisation flux density of $0.14\pm0.03$ \SI{}{\jansky}, resulting in a fractional polarisation of the source of around \SI{9.3}{\percent}. The peak polarisation lies around \SI{35}{\percent}.

Rotation measure (RM) maps are shown in Fig.~\ref{fig:askap-pi} above a threshold of \SI{6}{\SIsigma} (centre) and above \SI{10}{\SIsigma} (bottom), with pixels of PI intensity below \SI{90}{\micro\jansky\per\beam} and \SI{150}{\micro\jansky\per\beam} masked, respectively. Values for the RM range from $-150$ to $+\SI{350}{\radians\per\square\meter}$, with the maxima and minima seen overlapping with the brightest regions of the shell. The eastern shell shows a large positive RM, while the western shell indicates a very negative RM. The whole map (Fig.~\ref{fig:askap-pi}, centre) shows a gradient from positive RM towards the bottom left, to negative RM towards the top right. This gradient extends beyond the SNR and is likely related to the diffuse polarised emission observed throughout the image.

\subsection{Infrared (IR) results}
\begin{figure}
    \captionsetup[subfigure]{labelformat=empty}
    \centering
    \subfloat[\centering]{{\includegraphics[width=0.9\columnwidth, trim={1.1cm 0.0cm 1.5cm 1.6cm}, clip]{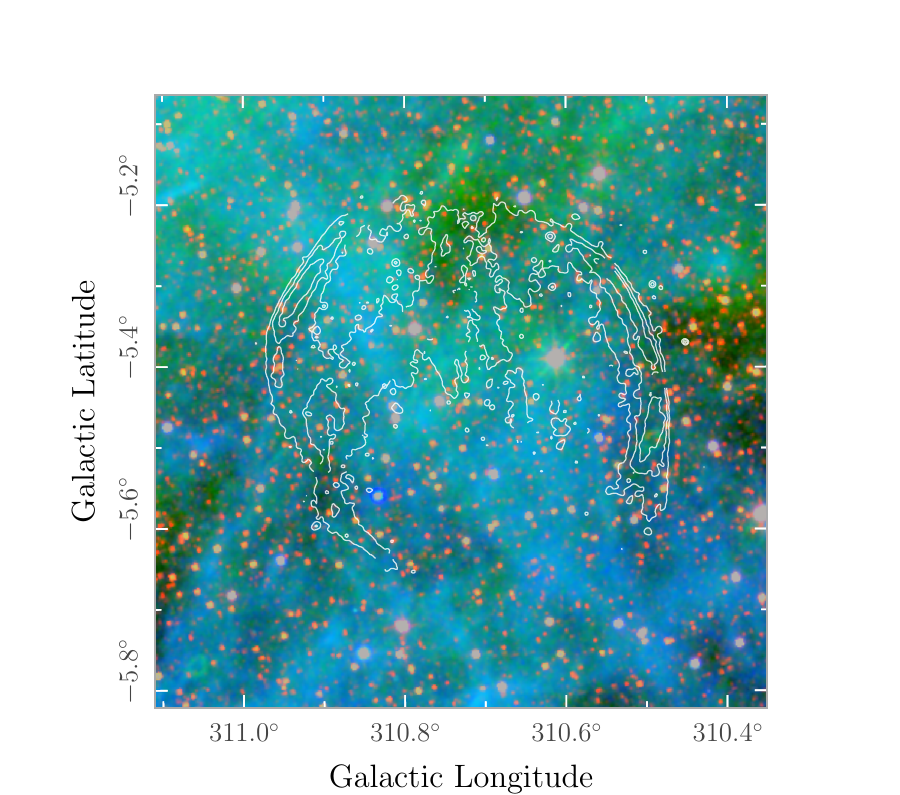}}}%
    \vspace{-0.4cm}
    \caption{RGB intensity image of WISE infrared (IR) data at 4.6 (red), 12 (green) and \SI{22}{\micro\m} (blue). White lines show total intensity contours in ASKAP. This RGB image was produced using methods described in~\citet{Lupton_2004}.}
    \label{fig:wise}
\end{figure}

We do not observe spatially correlated IR emission in WISE data that overlaps with the SNR shell. An RGB image of IR observations at 4.6, 12 and \SI{22}{\micro\m}, with overlaid ASKAP radio contours can be found in Fig.~\ref{fig:wise}.

Some filamentary structures can be seen at \SI{22}{\micro\meter}, and some excess towards the northern part of the shell in \SI{12}{\micro\meter}. Overall, though, there are no structures that follow the outline of the SNR in any way, which would be expected if the emission were indeed of thermal origin.

\subsection{Gamma-ray results}
In this study, we analysed existing sources within the 4FGL--DR4 \fermilat\ catalogue. Results from the analysis show the source positions, their spatial and spectral models, as well as their energy flux, and spectral parameters, together with their test statistic values in Table~\ref{tab:fermi-cat-sources}. Three sources in the vicinity of \snr\ SNR are shown. J1413 was modelled with a \ps\ spatial model and an \lp\ spectral model as taken from the catalogue.

\begin{figure*}
    \captionsetup[subfigure]{labelformat=empty}
    \subfloat[\centering]{{\includegraphics[width=1.05\columnwidth, trim={0.2cm 0cm 1cm 1.6cm}, clip]{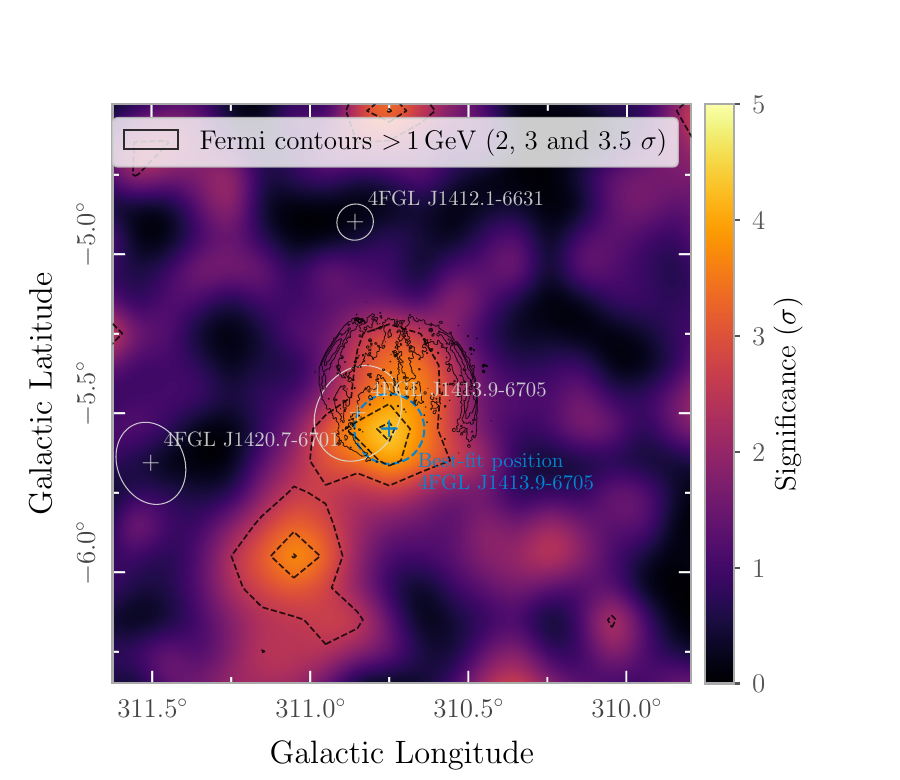}}}%
    \subfloat[\centering]{{\includegraphics[width=0.95\columnwidth, trim={0.0cm 0cm 0cm 0cm}, clip]{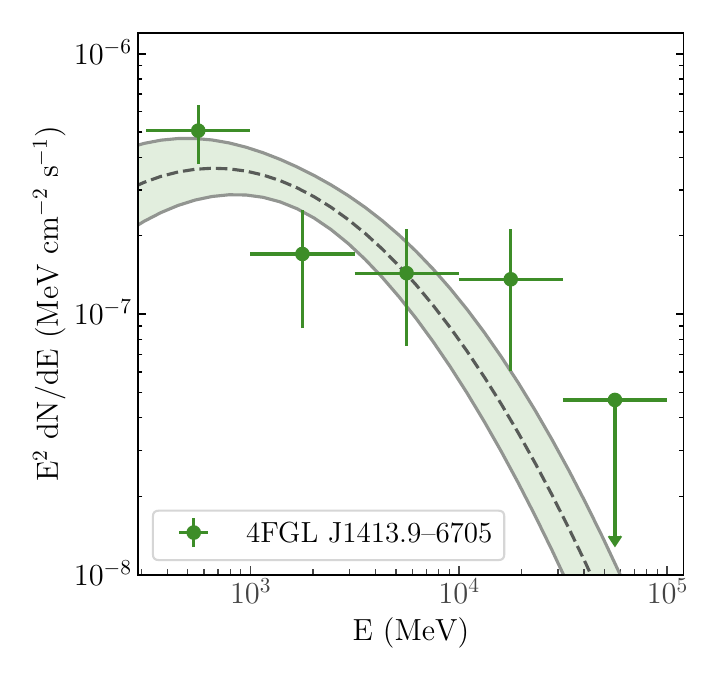}}}%
    \vspace{-0.4cm}
    \caption{\textbf{Left:} Significance maps of the \g\ region as seen with \fermilat\ with the optimised position of J1413 at $l,~b$ = \SI{310.75}{\degree}, \SI{-5.55}{\degree} (blue cross) in the energy range of \SI{100}{\MeV}--\SI{1}{\TeV}. The blue dashed circle indicates positional uncertainty for the new position of J1413. White circles indicate the \SI{68}{\percent} containment radii of respective sources. Thin black contours show radio excess, while dashed black lines indicate significance contours above \SI{1}{\GeV} (2, 3 and \SI{3.5}{\SIsigma} contours). \textbf{Right:} Spectrum of J1413 at the optimised source position, shown with statistical uncertainties. Four individual datapoints are presented together with one \SI{95}{\percent} upper limit at \SI{5.6e4}{\MeV}.}
    \label{fig:fermi}
\end{figure*}

In Fig.~\ref{fig:fermi} we present the results for the $\gamma$-ray analysis. The left panel shows the significance map of the \g\ region. While the other \fermilat\ catalogue sources are modelled, the emission attributed to J1413 is shown in the colourmap, and emphasised with significance contours at 2, 3 and \SI{3.5}{\SIsigma} levels. White crosses represent original catalogue source positions. The \texttt{localize()} method, which is used to optimise source positions by applying it to data above \SI{1}{\GeV}, placed J1413 at Galactic position $l, b$ = \SI{310.75}{\degree},~\SI{-5.55}{\degree}, slightly over \SI{0.1}{\degree} from the 4FGL catalogue position, marked by the blue cross and annotation. Positional fitting, despite the relatively large point-spread-function of \fermilat\ of \SI{0.45}{\degree} (\texttt{PSF3}) to \SI{0.88}{\degree} (\texttt{PSF1}) at \SI{1}{\GeV} (\SI{68}{\percent} containment)\footnote{\href{https://fermi.gsfc.nasa.gov/ssc/data/analysis/documentation/Cicerone/}{fermi.gsfc.nasa.gov/ssc/data/analysis/documentation/Cicerone/}}, is much more accurate, with a positional uncertainty (\SI{68}{\percent}) of \SI{0.1}{\degree} for the particular localisation results presented. In the 4FLG catalogue the \ps\ model had a \SI{68}{\percent} containment radius of $\SI{9.5}{\arcmin} \times \SI{7.8}{\arcmin}$. Other Fermi sources in the vicinity included 4FGL J1420.7--6701 and 4FGL J1412.1--6631 (to the left and above \snr\ in Fig.~\ref{fig:fermi} (left), respectively); their parameters are also shown in Table~\ref{tab:fermi-cat-sources}.

The map shows a significant maximum of emission coincident with the SNR with a significance of \SI{5.7}{\SIsigma}. The $\gamma$-ray emission overlaps with the radio emission well, although it does not align with its maximum. Overall, the positional match is very good, considering uncertainties in the $\gamma$-ray position and instrument resolution.

As results from spectral fitting were almost identical for the original and the optimised source position, we only show the spectral energy distribution (SED) of the new position in the right-hand panel of Fig.~\ref{fig:fermi}, with the corresponding SED bins and the energy flux presented in Table~\ref{tab:fermi-sed-values}. The best-fit spectral parameters for the whole energy range \SI{100}{\MeV} -- \SI{1}{\TeV} showed an energy flux of \SI{1.26\pm0.35e-6}{\MeV \per\cm\squared\per\s} (photon flux of \SI{2.62\pm1.39e-09}{\photon\per\cm\squared\per\s}) with a normalisation N$_0$ = \SI{7.12\pm1.64e-14}{\per\MeV\per\cm\squared\per\s}, $\alpha=2.45\pm0.25$, $\beta=0.21\pm0.12$, and E$_0=\SI{1997}{\MeV}$ (fixed parameter) for the source.

In total, four individual data points between $\sim\SI{0.6}{\GeV}$ and $\sim\SI{18}{\GeV}$, and one upper limit (\SI{95}{\percent}) at $\sim\SI{56}{\GeV}$ are shown in Fig.~\ref{fig:fermi} (right). Statistical uncertainties on energy and flux are plotted together with the respective data points.

\begin{table*}
    \renewcommand*{\arraystretch}{1}
        \centering
        \tiny
\caption{\label{tab:fermi-cat-sources} Fitting results of the three catalogued 4FGL--DR4 sources in the vicinity of \g\ using \SI{16.5}{\year} of \fermilat\ data, displayed together with their respective spatial and spectral models, as well as their respective positions. The main source overlapping with the SNR is emphasised in bold font.}
\begin{threeparttable}
    \begin{tabular}{lllllll}
    
    \hline\hline
    Source name & Position (GAL) & Models & Energy Flux & Spec. Index $\Gamma$ / $\alpha$  & $\beta$ &  TS \\
     &  &  & $\SI{100}{\MeV}-\SI{100}{\GeV} $ &  &  \\
     & (deg / deg)&  (spat./spec.) & (MeV cm$^{-2}$ s$^{-1} $) &  &  \\
    
    \hline
    \rule{0pt}{2ex}\!\!

    \textbf{4FGL~J1413.9--6705} & 310.75 / $-5.55$ & \pss$^{(*)}$/\lps$^{(\triangledown)}$ & \SI{1.26\pm0.35e-6}{} & $2.45\pm0.25$ & $0.21\pm0.12$ & \hspace{2px} 32.22\\
    4FGL~J1412.1--6631 & 310.86 / $-4.90$ & \pss/\lps$^{(\triangledown)}$ & \SI{3.54\pm0.33e-06}{} & $2.14\pm0.11$ & $0.49\pm0.10$ & 339.54 \\
    4FGL~J1420.7--6701 & 311.50 / $-5.66$ & \pss/\pls$^{(\dag)}$ & \SI{2.15\pm0.41e-6}{} & $2.42\pm0.13$ & --- & \hspace{2px} 41.34 \\
    \hline
    \end{tabular}

    \begin{tablenotes}
        \scriptsize
        \item[] \textbf{Notes.} ($* $) \ps\ spatial model; $(\triangledown)$ The log-parabola (LP) spectral model in \texttt{fermipy} is given as $\frac{\text{dN}}{\text{dE}} = \text{N}_0(\frac{\text{E}}{\text{E}_0})^{- \alpha - \beta \log(\text{E}/\text{E}_0)}$, with $\text{N}_0$ the normalisation or prefactor, $\alpha$ the spectral index, $\beta$ the curvature value, and \text{E}$_0$ the scaling energy; $(\dag)$ The power-law (PL) spectral model in \texttt{fermipy} is given as $\frac{\text{dN}}{\text{dE}} = \text{N}_0 (\frac{\text{E}}{\text{E}_0})^{-\Gamma}$, with N$_0$ the normalisation or prefactor, $\Gamma$ the spectral index, and E$_0$ the scaling energy;
    \end{tablenotes}
\end{threeparttable}
\end{table*}

\begin{table}
    \renewcommand*{\arraystretch}{1}
        \centering
\caption{\label{tab:fermi-sed-values} Spectral energy distribution (SED) parameters of J1413, corresponding to the data points shown in Fig.~\ref{fig:fermi} (right).}

\tiny
\begin{threeparttable}
    \begin{tabular}{rrrrr}
    
    \hline\hline
    E$_{\text{ctr}}$ & E$_{\text{min}}$ & E$_{\text{max}}$ & dN/dE & E$^2$dN/dE \\
    (MeV) & (MeV) & (MeV) & (MeV$^{-1}$ cm$^{-2}$ s$^{-1} $) & (MeV cm$^{-2}$ s$^{-1} $) \\
    
    \hline
    \rule{0pt}{2ex}\!\!
    
    \textbf{562} &  316 & 1000 & \SI{1.60e-12}{} & \SI{5.06e-07}{} \\
    \textbf{1778} & 1000 & 3162 & \SI{5.38e-14}{} & \SI{1.70e-07}{} \\
    \textbf{5623} & 3162 & \num{10000} & \SI{4.55e-15}{} & \SI{1.44e-07}{} \\
    \textbf{17\,783} & \num{10000} & \num{31622} & \SI{4.30e-16}{} & \SI{1.36e-07}{} \\
    
    \hline

    56\,234$^{(*)}$ & \num{31622} & \num{100000} & \SI{1.47e-17}{}$^{(*)}$ & \SI{4.68e-08}{}$^{(*)}$\\

    \hline
    \end{tabular}
    \begin{tablenotes}
        \scriptsize
        \item[] \textbf{Notes.} $(*)$ \SI{95}{\percent} confidence upper limit (UL);
    \end{tablenotes}
\end{threeparttable}
\end{table}

\section{Discussion}
\label{sec:discussion}

The discovered radio source \g\ shows typical radio morphology as seen in other bilateral shell SNRs, such as G296.5+10.0 and SN\,1006~\citep[e.g.][]{Giacani_2000AJ....119..281G, Roger_1988ApJ...332..940R, West_2016A&A...587A.148W}, with brighter regions in the north-east (top left) and west (right) in Fig.\ref{fig:askap-i}. The total intensity map (Fig.~\ref{fig:askap-i}) shows hints of a double shell, which could be explained by a density gradient around the remnant, with one side expanding into a denser medium than the other. As the ambient magneto-ionic medium shows both structure and field reversal, seen in the rotation measure (RM) gradient of the diffuse polarised emission (Fig.~\ref{fig:askap-pi}, centre), a conceivable assumption would be that the ambient density also shows structure as it is tied to the magnetic field. Simulations could give insight into the formation of such morphology, but would require assumptions on the local densities and magnetic fields.

There are strong indications that the emission is non-thermal: IR maps from WISE (Fig.~\ref{fig:wise}) show no correlation between the SNR's bilateral shell. The source is highly linearly polarised (Fig.~\ref{fig:askap-pi}, top) as expected from synchrotron emission, further supporting the non-thermal origin of the object.

\subsection{Distance}
\label{subsec:age}

Taking the flux density of $\sim\!\SI{1.5}{\jansky}$ at \SI{943.5}{\GHz}, as presented in Section \ref{sec:results-radio-continuum}, assuming a canonical spectral index of $-0.5$ for SNRs~\citep[i.e.][]{Blandford_1978}, and extrapolating the flux to \SI{1}{\GHz}, we calculated a radio surface brightness of $\Sigma_{\SI{1}{\GHz}} = 2.4^{+2.4}_{-0.1}\times10^{-22}\,\SI{}{\watt\per\square\meter\per\hertz\per\sterrad}$.
Despite the empirical $\Sigma-D$ relation~\citep{Shklovskii_1960} only being a crude estimate for the distance~\citep[e.g.][]{Green_2005}, we currently do not have other reasonable means to estimate the distance to \g. Using the updated SNR dataset from \citet{Pavlovic_2014}, we estimate the size of \g\ to be $42^{+42}_{-21}\,\SI{}{\parsec}$, with the uncertainties obtained from the quoted fractional uncertainty of a factor of 2 for an individual distance estimate using the $\Sigma-D$ relation \citep{Pavlovic_2014}. This results in a distance estimate of $4.9^{+4.9}_{-2.5}\,\SI{}{\kilo\parsec}$. Results from \citet{Green_2025JApA...46...14G} show that these uncertainties may be even larger.
The estimated distance would position \g\ around $460^{+460}_{-230}\,\SI{}{\parsec}$ below the Galactic plane. The distance estimate above seems to align well with methods developed in \citet{West_2016A&A...587A.148W} who modelled SNRs expanding into an ambient Galactic magnetic field, resulting in an estimate of $5-\SI{6}{\kilo\parsec}$~\citep{West_PhD_Thesis}.

The SNR has a diameter of $\sim\SI{30}{\arcmin}$, which is comparable to the angular scales where the ASKAP-EMU sensitivity to diffuse, large-scale emission begins to decline and drops to zero at $\sim43 - \SI{60}{\arcmin}$~\citep{Hopkins_2025PASA...42...71H}. Therefore, some flux may be missing in the observation. However, this is already reflected in the large intrinsic uncertainty given.
Even if the true flux (and hence surface brightness) were underestimated by a factor of 2, the resulting change in the $\Sigma-D$ distance would be less than \SI{15}{\percent} and would not alter our qualitative conclusions.

\subsection{Polarised emission}
\label{sec:discussion-polarised-emission}

Polarised emission is clearly seen in Fig.~\ref{fig:askap-pi}~(top) following the shape of the total intensity contours, and is especially prominent in the south-western and north-eastern regions where the total intensity is also strongest. The structure shows an area-averaged fractional polarisation of \SI{9.3}{\percent} over the whole remnant, well in line with expectations for non-thermal synchrotron emission. Assuming a spectral index of $-0.5$, one would expect a maximum degree of polarisation of \SI{69}{\percent}~\citep{Longair_2011hea..book.....L}, well above our results. Older remnants usually have a quite high fractional polarisation degree~\cite[e.g.][]{Kothes_1998A&A...331..661K}, while younger SNRs often show polarisation degrees below \SI{10}{\percent}~\cite[e.g.][for Cassiopeia A, Kepler, 1006, Tycho, respectively]{Anderson_1995ApJ...441..300A, DeLaney_2002ApJ...580..914D, Reynolds_1993AJ....106..272R, Dickel_1991AJ....101.2151D}. In these young shell-type SNRs, the reason for the lower fractional polarisation is the large amount of turbulence that leads to beam depolarisation. Another possibility for decreased fractional polarisation in more evolved SNRs is called depth depolarisation, which for an SNR would be internal depolarisation, caused by differential Faraday rotation \citep[e.g.][]{Burn_1966}.

By applying RM synthesis, we find that \snr\ SNR has a highly ordered RM structure, with the left shell displaying large positive values while the right is showing large negative regions. The gradient seen in Fig.~\ref{fig:askap-pi} (centre) could be related to a coinciding cloud, which would explain the amplification of the signal at the shell of the remnant with the same sign as the diffuse emission. Compared to large-scale magnetic field structures in \citet{Hutschenreuter_2022A&A...657A..43H} did not provide any additional information as the SNR is of the size of one pixel in the available RM maps.

From observing a regular SNR, expanding into a homogenous ambient medium, it would be expected that the RM structure of the remnant shells is entirely symmetric~\citep{Kothes_2009IAUS..259...75K}, i.e. showing the same sign in the rotation measure. Here, we find that one shell exhibits positive rotation measures, while the other shows negative rotation measures. This could mean that the SNR is expanding in a variable environment where the background polarised emission we observe originates from around the same distance as the SNR. Such a case could explain the gradient in the maps while permitting a Type~Ia thermonuclear explosion scenario that would also more easily align with \snr's high-latitude position.

One may also interpret the observed RM signal as evidence that the remnant expands in a toroidal magnetic field~\citep{Harvey-Smith_2010ApJ...712.1157H}, implying that the SNR is expanding in a stellar wind bubble of the progenitor, creating the need for a massive progenitor star and a Type~II supernova (SN) explosion in this case. Although this core-collapse scenario is possible, it does face significant difficulties.

\subsection{Abeona's multiwavelength context}

Results published for MOST and MWA show lower limits for the total flux density of $>\SI{0.83}{\jansky}$ at \SI{843}{\mega\hertz}~\citep{GreenA_2014} and $>\SI{1.1}{\jansky}$ at \SI{200}{\MHz}~\citep{Mantovanini_2025PASA...42...21M}, respectively, which is consistent with measurements presented in this work, assuming the synchrotron emission follows a powerlaw with a spectral index of $-0.5$. However, no meaningful constraints can be deduced from these lower limits, except that the emission is optically thin.
The lack of correlating IR emission is also consistent with the SNR interpretation of the source.

Searching publicly available archival data of \textit{ROSAT} and \textit{eROSITA} around the position of \g, we did not find any X-ray emission in either dataset. In most young SNRs (such as Cas A, Tycho and Kepler) X-ray synchrotron emission can be seen, but only in a few, such as SN\,1006, X-ray synchrotron emission is the dominant form of X-ray emission~\citep{Ballet_2006AdSpR..37.1902B}. It is rather unlikely that \snr\ is actually a young object, due to its low-surface brightness. The low-density ambient medium at such a Galactic latitude, as well as the possibly large distance, could both contribute to a low probability of detecting X-ray signals with current instruments.

We found a high-energy $\gamma$-ray source spatially coinciding with \g\ that could be connected to the remnant. With an energy flux of \SI{1.26\pm0.35e-6}{\MeV \per\cm\squared\per\s} (\SI{2.02\pm0.56e-12}{\erg \per\cm\squared\per\s}) and very low photon flux, it is not possible to determine whether the $\gamma$-rays are of leptonic or hadronic origin. However, all confirmed hadronic $\gamma$-ray emissions have been seen in the Galactic plane where SNRs interact with molecular clouds to accelerate protons that then produce $\gamma$-rays via the decay of neutral pions. Since \snr\ is most likely expanding in a very low-density environment, and no over-densities are apparent in CO maps 
from \textit{Planck}, the leptonic inverse-Compton scattering process, caused by the propagation of highly relativistic electrons in magnetic fields, is considered the more likely scenario.

\subsection{Abeona in context with other High-Latitude SNRs with $\gamma$-ray counterparts}

The diffuse emission of the Galactic plane and the risk of source confusion make it challenging to confidently detect SNRs, especially at GeV and TeV energies and within a one-degree radius of the plane. This, in turn, makes studying particle escape and detecting flux gradients at the edge of an SNR particularly difficult. To reveal the physics of CR escape from SNRs and how they contribute to the Galactic CR population, observations outside the Galactic plane are needed.

\g\ is one of a few SNRs that have been discovered displaying $\gamma$-ray emission at high Galactic latitudes. In \citet{Burger-Scheidlin_2024}, we compiled a list of such sources together with their spatial and spectral $\gamma$-ray properties. \citet{Araya_2024}, in addition to detecting another such source, suggested a few more SNRs fitting this subset. Therefore, in Table~\ref{tab:highlat_sources}, we present an updated list of these high Galactic SNRs visible at $\gamma$-ray energies at Galactic latitudes more than \SI{4}{\degree} above or below the Galactic plane, and have expanded the table to also show their observational radio properties.

Results in Table~\ref{tab:highlat_sources} are sorted by their position in Galactic longitude. Both in terms of radio flux, as well as $\gamma$-ray flux, \g\ is one of the faintest objects on the list. It is one of five objects modelled with an \lp\ model and shows one of the softest $\gamma$-ray spectra.

Overall, the SNRs range from less than a tenth of a degree in radio extension, up to more than \SI{1.7}{\degree} in radius. Compared to their $\gamma$-ray counterparts, no clear correlation can be seen, as some of them are larger than in radio frequencies, while others appear point-like. Considering CR diffusion in SNRs, \citet{Brose_2021A&A...654A.139B} predict that older SNRs may appear larger at $\gamma$-ray energies compared to radio and exhibit halos.
Radio fluxes at \SI{1}{\GHz} range between \SI{0.29}{\jansky} all the way to more than \SI{200}{\jansky}. Due to the potentially different nature of emission and expected different magnetic field strength in these objects, correlation of the $\gamma$-ray and radio emission is possible but mostly not expected. From this dataset, no clear correlation is evident.
Considering the radio spectral indices, a range between $-0.38$ and $-0.75$ is observed, centred around the expected value of $-0.5$ from diffusive shock acceleration theory. The same SNRs in $\gamma$-rays show values for the spectral index range between 1.5 for hard sources, all the way to 2.7 for softer sources, with a log-parabola spectral model being a better description of about a third of them. 

From the total sample of 300 confirmed SNRs in the Galaxy, fewer than 40 are detected at GeV\,--\,TeV energies. Out of these 40, around one third fall into this sample of objects at high Galactic latitudes presented in this work. SNRs like these would likely be much harder to detect in the plane of the Galaxy due to the diffuse emission that can easily obscure low-surface-brightness emission.

\section{Conclusions}
\label{sec:conclusion}

In this work, we present recent ASKAP radio observations of a newly confirmed SNR found at high Galactic latitude. Radio continuum data at \SI{943.5}{\MHz} show the object with a symmetric circular, bilateral shell morphology. No infrared counterparts can be seen in WISE infrared data, strongly suggesting non-thermal emission. Radio polarisation maps demonstrate that significant linearly polarised emission can be seen from the bilateral shell of the object, characteristic of synchrotron emission. Based on these observations, we conclude that Abeona / \g\ is a newly discovered SNR. Additionally, a spatially overlapping $\gamma$-ray source suggests that the SNR could be accelerating particles to high energies.
Considering its Galactic position, lack of identified compact-object remnant, \g's precursor was most likely a Type~Ia supernova explosion.

Abeona is now the 13th object of a subset of SNRs off the Galactic plane showing significant high-energy emission. SNRs at these high Galactic latitudes are ideal for testing cosmic ray acceleration and diffusion.

Additional radio flux points would allow for spectral index estimation, which instruments like \textit{MeerKAT} could potentially provide. The upcoming \textit{Square Kilometre Array Observatory} (SKAO) will, in its fully operational state, have the ability to detect low-surface brightness objects, such as this source.
From a multiwavelength perspective, an analysis of up-to-date eROSITA data, as well as pointed XMM-Newton and Athena observations, could be very useful and would raise interesting questions in case X-rays are detected. The upcoming \textit{Cherenkov Telescope Array Observatory} (CTAO) could provide the sensitivity and resolution needed to detect this remnant at very high energies above some tens of GeV. It could be possible to detect Abeona with deep optical observations, similar to Nereides SNR.

\begin{landscape}
    \topskip0pt
    \vspace*{\fill}
   
    \begin{threeparttable}[htb!]

    \caption{Comparison of \g\ to fluxes and photon spectral indices of other known high-latitude SNRs ($\,|\,\text{b}\,|>\SI{4}{\degree}$) detected at high energies, sorted by their Galactic longitude. The list includes data from the SNR catalogue provided by \citet{Ferrand_2012}.$^{(\diamond)}$ As suggested by~\citet{Araya_2024}, additional such sources were included in the table. Photon spectral indices provided indicate the use of a power-law spectral model unless a value for $\beta$ is given, indicating a log-parabola spectral model.}
    \label{tab:highlat_sources}

        \centering
        \tiny
        \setlength{\tabcolsep}{2pt}
        \begin{tabular}{l|llll|llllll}
        \hline
        \hline
         & \multicolumn{4}{c}{Radio} & \multicolumn{5}{c}{Gamma-ray} \\
         & \multicolumn{4}{c}{\downbracefill} & \multicolumn{5}{c}{\downbracefill}\\ 
         
        Source Name & Extension & Flux & Spectral index $\alpha$ & Reference & Extension & \multicolumn{1}{l}{Energy Flux} &  Spectral Index $\Gamma$ / $\alpha$ & \hspace{0.1cm}$\beta$ & Reference\\
         & (deg) & (Jy) & --- &  & (deg) & \multicolumn{1}{l}{(\SI{}{\mega\electronvolt\per\square\cm\per\s})} & --- & \hspace{0.1cm}--- & \\
         &  & \SI{1}{\GHz} &  &  &  & \SI{1}{GeV} -- \SI{1}{\TeV} & \\
        \hline

        \makecell[l]{G4.5+6.8 \\ \hspace{7px} \tiny Kepler SNR \\ \hspace{7px} \tiny SN\,1604} & 0.06 & $18.5\pm0.3$\tnote{$(\bot)$} & $-0.71\pm0.2$ & \cite{DeLaney_2002ApJ...580..914D} & PS$^{(*)}$ & $1.21^{+1.8}_{-0.6}\times\SI{e-6}{}$\tnote{$(\bot)$} & $2.14 \pm 0.12_{\text{stat}} \pm 0.15_{\text{sys}}$ & --- &  \cite{Acero_2022} \\
        \hdashline[0.75pt/1pt]
        
        G17.8+16.7 & $\sim 0.8$ & $2.7\pm0.1$\tnote{$(\bot)$} & $-0.75\pm0.15$ & \cite{Araya_2022} & \SI{0.73}{} & \makecell[l]{\hfill \\ \SI{1.38\pm0.26e-05}{}\tnote{$(\dag)$}} & \makecell[l]{$1.83\pm0.02_{\text{stat}}\pm 0.05_{\text{sys}}$ \\ $1.97\pm0.08_{\text{stat}}\pm0.06_{\text{sys}}$} & --- & \makecell[l]{\cite{Araya_2022} \\ \cite{Ackermann_2018}}\\
        \hdashline[0.75pt/1pt]

        \makecell[l]{G74.0$-$8.5 \\ \hspace{7px} \tiny Cygnus Loop}  & $3.5\times2.5$ & $168\pm15$\tnote{$(\bot)$} & $-0.42\pm0.06$ & \makecell[l]{\cite{2004AA...426..909U} \\ \cite{2006AA...447..937S}} & $1.6\pm0.1$$^{(\bot\bot)}$ & $4.1^{+0.5}_{-0.4}\times\SI{e-5}{}$\tnote{($\ddag\ddag$)} &  $2.02\pm0.03$ & $0.27\pm0.02$ & \cite{Katagiri_2011} \\
        \hdashline[0.75pt/1pt]

        \makecell[l]{G89.0+4.7 \\ \hspace{7px} \tiny HB\,21} & $2.5\times2.3$ & $194\pm19$\tnote{$(\bot)$} & $-0.36\pm0.03$ & \makecell[l]{\cite{2006_Kothes} \\ \cite{Gao_2011AA...529A.159G}}& $1.19\pm0.06$ & \SI{5.87\pm1.5e-5}{}$^{(\triangledown\triangledown)}$ & $2.54\pm0.05$ & $0.39 \pm 0.04$ & \cite{Pivato_2013} \\
        \hdashline[0.75pt/1pt]
        
        \makecell[l]{G107.7$-$5.1 \\ \hspace{7px} \tiny Nereides SNR} & & no radio detection &  & \cite{Fesen_2024} & $1.02 \pm 0.03$ & \SI{5.8e-06}{} & $1.67 \pm 0.01_{\text{stat}} \pm 0.12_{\text{sys}}$ & $0.22 \pm 0.01_{\text{stat}}\text{\,}^{+0.09}_{-0.04} \text{}_\text{sys}$ & \cite{Araya_2024} \\
        \hdashline[0.75pt/1pt]

        \makecell[l]{G118.4+37.0 \\ \hspace{7px} \tiny Calvera SNR} & $\sim 0.95$ & $0.3\pm0.2$\tnote{$(\bot)$} & $-0.71\pm0.09$ & \cite{Arias_2022} & \SI{0.53}{}  & \SI{3.06e-06}{}\tnote{$(\bot)$} & $1.66\pm0.10_{\text{stat}}\pm 0.03_{\text{sys}}$ & --- & \cite{Araya_2023} \\
        \hdashline[0.75pt/1pt]
        
        G150.3+4.5 & $2.5\times3$ & --- & $-0.4$ to $-0.7$\tnote{$(\triangledown)$} & \makecell[l]{\cite{Gerbrandt_2014AA...566A..76G} \\ \cite{Gao_2014AA...567A..59G}} & \SI{1.5}{}\tnote{($\bowtie$)} & \SI{5.20e-05}{}\tnote{$(\bot)$} & $1.62\pm0.04_{\text{stat}}\pm 0.22_{\text{sys}}$ & $0.07\pm 0.02_{\text{stat}}\pm 0.02_{\text{sys}}$ & \cite{Devin_2020} \\
        \hdashline[0.75pt/1pt]

        G166+4.3 & $1.6\times0.8$ & $5.8\pm0.4$\tnote{$(\bot)$} & $-0.37\pm0.11$ & \cite{2006_Kothes} & $\sim\SI{0.3}{}$ & \SI{2.87e-06}{}\tnote{$(\bot)$} & $2.7\,\,\,\pm0.1$ & --- & \cite{Araya_2013} \\  
        \hdashline[0.75pt/1pt]

        \makecell[l]{G288.8$-$6.3 \\ \hspace{7px} \tiny Ancora SNR} & $\sim0.85$ & $10.9\pm0.1$ & $-0.41\pm0.12$ & \cite{Filipovic_2023} & $0.92\pm0.07$ & \SI{3.29\pm0.78e-6}{} &  $2.31\pm0.11$\tnote{$(\bot)$} & --- & \cite{Burger-Scheidlin_2024} \\
        \hdashline[0.75pt/1pt]

        G296.5+10.0 & $1.6\times1.1$ & $51\pm6$\tnote{$(\bot)$} & --0.5 & \makecell[l]{\cite{1994MNRAS.270..106M} \\ \cite{Cotton_2024ApJS..270...21C}} & \SI{0.7}{} & \makecell[l]{ \SI{8.17e-06}{}\tnote{$(\bot)$} \\ \SI{1.13\pm0.24e-05}{}\tnote{$(\dag)$}} & \makecell[l]{$1.85\pm0.13$ \\ $1.81\pm0.09_{\text{stat}}\pm0.05_{\text{sys}}$ }& --- & \makecell[l]{\cite{Araya_2013} \\ \cite{Ackermann_2018}}\\
        \hdashline[0.75pt/1pt]

        \makecell[l]{\textbf{\g} \\ \hspace{7px} \tiny \textbf{\snr\ SNR}} & 0.25 & $1.5^{+1.5}_{-0.1}$ & -- ? -- & \textbf{This work} & PS$^{(*)}$ & \SI{5.56\pm1.36e-7}{} &  $2.45\pm0.25$ &  $0.21\pm0.12$ & \textbf{This work} \\
        \hdashline[0.75pt/1pt]
        
        \makecell[l]{G327.6+14.6 \\ \hspace{7px} \tiny SN\,1006} & $\sim 0.5$ & $22.3\pm0.3$\tnote{$(\bot)$} & $-0.52\pm0.02$ & \makecell[l]{\cite{2009AJ....137.2956D} \\ \cite{2024ApJ...970L..27T}} & 0.1 & \SI{3.63\pm1.62e-06}{}\tnote{(**)} & $1.57\pm0.11$ & --- & \cite{Condon_2017}\\
        \hdashline[0.75pt/1pt]
        
        G332.5$-$5.6  & $\sim 0.5$ & $2.4\pm0.2$\tnote{$(\bot)$} & $-0.7\pm0.2$ & \cite{2007MNRAS.375...92R} & 0.54\tnote{($\ddag$)} & $\sim$ \SI{8.8e-6}{}\tnote{$(\bot)$} &  $2.14\pm0.04$ & --- & \cite{Luo_2024} \\
                
        \hline
        \end{tabular}

        \begin{tablenotes}
        \scriptsize
        \item[] \textbf{Notes.} ($\diamond$) \href{http://snrcat.physics.umanitoba.ca}{snrcat.physics.umanitoba.ca}; (*) \ps\ spatial model, the source extension is smaller than the PSF of \fermilat; ($\bowtie$) using results for the radial Gaussian model; ($\bot$) calculated using data provided in the respective publication; ($\triangledown$) $-0.4$ for the eastern shell, $-0.7$ for the western shell; (\dag) from FITS data provided with~\citet{Ackermann_2018}; (**) range is $\SI{1}{\GeV} - \SI{2}{\TeV}$;  ($\bot\bot$) ring shape, outer ring radius given in Table, inner ring radius $0.7\pm0.1$; ($\ddag$) using the radial disk model; ($\ddag\ddag$) range is $0.2-\SI{100}{\GeV}$;  ($\triangledown\triangledown$) range is $0.1-\SI{300}{\GeV}$;

    \end{tablenotes}

    \end{threeparttable}
    
    \vspace*{\fill}
\end{landscape}


\begin{acknowledgements}
    This scientific work uses data obtained from Inyarrimanha Ilgari Bundara / the Murchison Radio-astronomy Observatory. We acknowledge the Wajarri Yamaji People as the Traditional Owners and native title holders of the Observatory site. The Australian Commonwealth Scientific and Industrial Research Organisation's (CSIRO) ASKAP radio telescope is part of the Australia Telescope National Facility\footnote{\href{https://ror.org/05qajvd42}{ror.org/05qajvd42}}. Operation of ASKAP is funded by the Australian Government with support from the National Collaborative Research Infrastructure Strategy. ASKAP uses the resources of the Pawsey Supercomputing
    Research Centre. Establishment of ASKAP, Inyarrimanha Ilgari Bundara, the CSIRO Murchison Radio-astronomy Observatory, and the Pawsey Supercomputing Research Centre are initiatives of the Australian Government, with support from the Government of Western Australia and the Science and Industry Endowment Fund. The POSSUM project\footnote{\href{https://possum-survey.org}{possum-survey.org}} has been made possible through funding from the Australian Research Council, the Natural Sciences and Engineering Research Council of Canada, the Canada Research Chairs Program, and the Canada Foundation for Innovation. \\

    C.B.S. acknowledges support from a Royal Society-Science Foundation Ireland Research Fellows Enhancement Award 2021 (22/RS-EA/3810). 
    
    J.M. acknowledges support from a Royal Society -- Science Foundation Ireland University Research Fellowship (20/RS-URF-R/3712).
    
    This publication results from research conducted with the financial support of Taighde \'Eireann - Research Ireland under Grant numbers 20/RS-URF-R/3712 and 22/RS-EA/3810.

    B.B. and R.K. acknowledge the support of the Natural Sciences and Engineering Research Council of Canada (NSERC), funding reference numbers RGPIN-2022-03499.

    S.L. and M.D.F. acknowledge Australian Research Council (ARC) funding through grant DP200100784. 
    
    This work makes use of publicly available \textit{Fermi}--LAT data provided online by the Fermi Science Support Center\footnote{\href{http://fermi.gsfc.nasa.gov/ssc/}{fermi.gsfc.nasa.gov/ssc/}}, as well as publicly available data obtained with \textit{Planck}\footnote{\href{http://www.esa.int/Planck}{esa.int/Planck}}, an ESA science mission. This research made use of \textit{Fermitools}\footnote{\href{https://github.com/fermi-lat/Fermitools-conda}{github.com/fermi-lat/Fermitools-conda}}, as well as the following Python packages: \textit{Fermipy}~\citep{Atwood_2013}, \textit{Astropy}~\citep{Astropy_2022}, \textit{Numpy}~\citep{Numpy_2020}, and \textit{Matplotlib}~\citep{Hunter_2007}. \\
    
    This publication makes use of data products from the Wide-field Infrared Survey Explorer, which is a joint project of the University of California, Los Angeles, and the Jet Propulsion Laboratory/California Institute of Technology, funded by the National Aeronautics and Space Administration.
    Access to the scripts used to produce this work will be made available upon contact with the corresponding author.
    
\end{acknowledgements}

\vspace{4cm}

%
  \bibliographystyle{aa_url} 
  \bibliography{bibliography.bib} 

@article{Hess_1912,
    author = {Hess, V. F.},
    title = {Über Beobachtungen der durchdringenden Strahlung bei sieben Freiballonfahrten},
    year = 1912,
    journal = {Physikalische Zeitschrift},
    volume = {13},
    pages = {1084-1091},
    }

@article{
Baade_and_Zwicky_1934,
author = {W. Baade  and F. Zwicky },
title = {Cosmic Rays from Super-Novae},
journal = {Proceedings of the National Academy of Sciences},
volume = {20},
number = {5},
pages = {259-263},
year = {1934},
doi = {10.1073/pnas.20.5.259},
URL = {https://www.pnas.org/doi/abs/10.1073/pnas.20.5.259},
eprint = {https://www.pnas.org/doi/pdf/10.1073/pnas.20.5.259}}

@ARTICLE{Hurley-Walker_2019PASA...36...48H,
       author = {{Hurley-Walker}, N. and {Gaensler}, B.~M. and {Leahy}, D.~A. and {Filipovi{\'c}}, M.~D. and {Hancock}, P.~J. and {Franzen}, T.~M.~O. and {Offringa}, A.~R. and {Callingham}, J.~R. and {Hindson}, L. and {Wu}, C. and {Bell}, M.~E. and {For}, B. -Q. and {Johnston-Hollitt}, M. and {Kapi{\'n}ska}, A.~D. and {Morgan}, J. and {Murphy}, T. and {McKinley}, B. and {Procopio}, P. and {Staveley-Smith}, L. and {Wayth}, R.~B. and {Zheng}, Q.},
        title = "{Candidate radio supernova remnants observed by the GLEAM survey over 345{\textdegree} < l < 60{\textdegree} and 180{\textdegree} < l < 240{\textdegree}}",
      journal = {PASA},
         year = 2019,
        month = nov,
       volume = {36},
          eid = {e048},
        pages = {e048},
          doi = {10.1017/pasa.2019.33},
archivePrefix = {arXiv},
       eprint = {1911.08124},
 primaryClass = {astro-ph.HE},
       adsurl = {https://ui.adsabs.harvard.edu/abs/2019PASA...36...48H}
}

@ARTICLE{Hurley-Walker_2019PASA...36...45H,
       author = {{Hurley-Walker}, N. and {Filipovi{\'c}}, M.~D. and {Gaensler}, B.~M. and {Leahy}, D.~A. and {Hancock}, P.~J. and {Franzen}, T.~M.~O. and {Offringa}, A.~R. and {Callingham}, J.~R. and {Hindson}, L. and {Wu}, C. and {Bell}, M.~E. and {For}, B. -Q. and {Johnston-Hollitt}, M. and {Kapi{\'n}ska}, A.~D. and {Morgan}, J. and {Murphy}, T. and {McKinley}, B. and {Procopio}, P. and {Staveley-Smith}, L. and {Wayth}, R.~B. and {Zheng}, Q.},
        title = "{New candidate radio supernova remnants detected in the GLEAM survey over 345{\textdegree} < l < 60{\textdegree}, 180{\textdegree} < l < 240{\textdegree}}",
      journal = {PASA},
         year = 2019,
        month = nov,
       volume = {36},
          eid = {e045},
        pages = {e045},
          doi = {10.1017/pasa.2019.34},
archivePrefix = {arXiv},
       eprint = {1911.08126},
 primaryClass = {astro-ph.HE},
       adsurl = {https://ui.adsabs.harvard.edu/abs/2019PASA...36...45H}
}

@ARTICLE{Smeaton_2024MNRAS.534.2918S,
       author = {{Smeaton}, Zachary J. and {Filipovi{\'c}}, Miroslav D. and {Lazarevi{\'c}}, Sanja and {Alsaberi}, Rami Z.~E. and {Ahmad}, Adeel and {Araya}, Miguel and {Ball}, Brianna D. and {Bordiu}, Cristobal and {Buemi}, Carla S. and {Bufano}, Filomena and {Dai}, Shi and {Haberl}, Frank and {Hopkins}, Andrew M. and {Ingallinera}, Adriano and {Jarrett}, Thomas and {Koribalski}, B{\"a}rbel S. and {Kothes}, Roland and {Kraan-Korteweg}, Ren{\'e}e C. and {Leahy}, Denis and {Lundqvist}, Peter and {Maitra}, Chandreyee and {Martin}, Pierrick and {Payne}, Jeffrey L. and {Rowell}, Gavin and {Sano}, Hidetoshi and {Sasaki}, Manami and {Soria}, Roberto and {Steyn}, Nadia and {Umana}, Grazia and {Uro{\v{s}}evi{\'c}}, Dejan and {Velovi{\'c}}, Velibor and {Vernstrom}, Tessa and {Vukoti{\'c}}, Branislav and {West}, Jennifer},
        title = "{Discovery of Perun (G329.9-0.5): a new, young, Galactic SNR}",
      journal = {\mnras},
         year = 2024,
        month = nov,
       volume = {534},
       number = {3},
        pages = {2918-2937},
          doi = {10.1093/mnras/stae2237},
       adsurl = {https://ui.adsabs.harvard.edu/abs/2024MNRAS.534.2918S}
}

@ARTICLE{Filipovic_2022MNRAS.512..265F,
       author = {{Filipovi{\'c}}, Miroslav D. and {Payne}, J.~L. and {Alsaberi}, R.~Z.~E. and {Norris}, R.~P. and {Macgregor}, P.~J. and {Rudnick}, L. and {Koribalski}, B.~S. and {Leahy}, D. and {Ducci}, L. and {Kothes}, R. and {Andernach}, H. and {Barnes}, L. and {Boji{\v{c}}i{\'c}}, I.~S. and {Bozzetto}, L.~M. and {Brose}, R. and {Collier}, J.~D. and {Crawford}, E.~J. and {Crocker}, R.~M. and {Dai}, S. and {Galvin}, T.~J. and {Haberl}, F. and {Heber}, U. and {Hill}, T. and {Hopkins}, A.~M. and {Hurley-Walker}, N. and {Ingallinera}, A. and {Jarrett}, T. and {Kavanagh}, P.~J. and {Lenc}, E. and {Luken}, K.~J. and {Mackey}, D. and {Manojlovi{\'c}}, P. and {Maggi}, P. and {Maitra}, C. and {Pennock}, C.~M. and {Points}, S. and {Riggi}, S. and {Rowell}, G. and {Safi-Harb}, S. and {Sano}, H. and {Sasaki}, M. and {Shabala}, S. and {Stevens}, J. and {van Loon}, J. Th and {Tothill}, N.~F.~H. and {Umana}, G. and {Uro{\v{s}}evi{\'c}}, D. and {Velovi{\'c}}, V. and {Vernstrom}, T. and {West}, J.~L. and {Wan}, Z.},
        title = "{Mysterious odd radio circle near the large magellanic cloud - an intergalactic supernova remnant?}",
      journal = {\mnras},
         year = 2022,
        month = may,
       volume = {512},
       number = {1},
        pages = {265-284},
          doi = {10.1093/mnras/stac210},
archivePrefix = {arXiv},
       eprint = {2201.10026},
 primaryClass = {astro-ph.HE},
       adsurl = {https://ui.adsabs.harvard.edu/abs/2022MNRAS.512..265F}
}

@ARTICLE{Aloisio_2012APh....39..129A,
       author = {{Aloisio}, R. and {Berezinsky}, V. and {Gazizov}, A.},
        title = "{Transition from galactic to extragalactic cosmic rays}",
      journal = {Astroparticle Physics},
         year = 2012,
        month = dec,
       volume = {39},
        pages = {129-143},
          doi = {10.1016/j.astropartphys.2012.09.007},
archivePrefix = {arXiv},
       eprint = {1211.0494},
 primaryClass = {astro-ph.HE},
       adsurl = {https://ui.adsabs.harvard.edu/abs/2012APh....39..129A}
}

@ARTICLE{Globus_2015PhRvD..92b1302G,
       author = {{Globus}, Noemie and {Allard}, Denis and {Parizot}, Etienne},
        title = "{A complete model of the cosmic ray spectrum and composition across the Galactic to extragalactic transition}",
      journal = {\prd},
         year = 2015,
        month = jul,
       volume = {92},
       number = {2},
          eid = {021302},
        pages = {021302},
          doi = {10.1103/PhysRevD.92.021302},
archivePrefix = {arXiv},
       eprint = {1505.01377},
 primaryClass = {astro-ph.HE},
       adsurl = {https://ui.adsabs.harvard.edu/abs/2015PhRvD..92b1302G}
}

@ARTICLE{Burn_1966,
       author = {{Burn}, B.~J.},
        title = "{On the depolarization of discrete radio sources by Faraday dispersion}",
      journal = {\mnras},
         year = 1966,
        month = jan,
       volume = {133},
        pages = {67},
          doi = {10.1093/mnras/133.1.67},
       adsurl = {https://ui.adsabs.harvard.edu/abs/1966MNRAS.133...67B}
}

@ARTICLE{Filipovic_2023,
       author = {{Filipovi{\'c}}, Miroslav D. and {Dai}, Shi and {Arbutina}, Bojan and {Hurley-Walker}, Natasha and {Brose}, Robert and {Becker}, Werner and {Sano}, Hidetoshi and {Uro{\v{s}}evi{\'c}}, Dejan and {Jarrett}, T.~H. and {Hopkins}, Andrew M. and {Alsaberi}, Rami Z.~E. and {Alsulami}, R. and {Bordiu}, Cristobal and {Ball}, Brianna and {Bufano}, Filomena and {Burger-Scheidlin}, Christopher and {Crawford}, Evan and {English}, Jayanne and {Haberl}, Frank and {Ingallinera}, Adriano and {Kapinska}, Anna D. and {Kavanagh}, Patrick J. and {Koribalski}, B{\"a}rbel S. and {Kothes}, Roland and {Lazarevi{\'c}}, Sanja and {Mackey}, Jonathan and {Rowell}, Gavin and {Leahy}, Denis and {Loru}, Sara and {Macgregor}, Peter J. and {Nicastro}, Luciano and {Norris}, Ray P. and {Riggi}, Simone and {Sasaki}, Manami and {Stupar}, Milorad and {Trigilio}, Corrado and {Umana}, Grazia and {Vernstrom}, Tessa and {Vukoti{\'c}}, Branislav},
        title = "{EMU Detection of a Large and Low Surface Brightness Galactic SNR G288.8-6.3}",
      journal = {\aj},
         year = 2023,
        month = oct,
       volume = {166},
       number = {4},
          eid = {149},
        pages = {149},
          doi = {10.3847/1538-3881/acf19c},
archivePrefix = {arXiv},
       eprint = {2308.08716},
 primaryClass = {astro-ph.HE},
       adsurl = {https://ui.adsabs.harvard.edu/abs/2023AJ....166..149F}
}

@ARTICLE{Burger-Scheidlin_2024,
       author = {{Burger-Scheidlin}, Christopher and {Brose}, Robert and {Mackey}, Jonathan and {Filipovi{\'c}}, Miroslav D. and {Goswami}, Pranjupriya and {Guillen}, Enrique Mestre and {de O{\~n}a Wilhelmi}, Emma and {Sushch}, Iurii},
        title = "{Gamma-ray detection of newly discovered Ancora supernova remnant: G288.8-6.3}",
      journal = {\aap},
         year = 2024,
        month = apr,
       volume = {684},
          eid = {A150},
        pages = {A150},
          doi = {10.1051/0004-6361/202348348},
archivePrefix = {arXiv},
       eprint = {2310.14431},
 primaryClass = {astro-ph.HE},
       adsurl = {https://ui.adsabs.harvard.edu/abs/2024A&A...684A.150B}
}

@ARTICLE{Araya_2024,
       author = {{Araya}, Miguel},
        title = "{Nonthermal GeV emission from the Nereides nebula: Confirming the nature of the supernova remnant G107.7‑5.1}",
      journal = {\aap},
         year = 2024,
        month = nov,
       volume = {691},
          eid = {A225},
        pages = {A225},
          doi = {10.1051/0004-6361/202451443},
archivePrefix = {arXiv},
       eprint = {2409.14006},
 primaryClass = {astro-ph.HE},
       adsurl = {https://ui.adsabs.harvard.edu/abs/2024A&A...691A.225A}
}

@ARTICLE{2025arXiv250504041F,
       author = {{Filipovic}, Miroslav D. and {Smeaton}, Zachary J. and {Kothes}, Roland and {Mantovanini}, Silvia and {Kostic}, Petar and {Leahy}, Denis and {Ahmad}, Adeel and {Anderson}, Gemma E. and {Araya}, Miguel and {Ball}, Brianna and {Becker}, Werner and {Bordiu}, Cristobal and {Bradley}, Aaron C. and {Brose}, Robert and {Burger-Scheidlin}, Christopher and {Dai}, Shi and {Duchesne}, Stefan and {Galvin}, Timothy J. and {Hopkins}, Andrew M. and {Hurley-Walker}, Natasha and {Koribalski}, Barbel S. and {Lazarevic}, Sanja and {Lundqvist}, Peter and {Mackey}, Jonathan and {Martin}, Pierrick and {McGee}, Padric and {Mitrasinovic}, Ana and {Payne}, Jeffrey L. and {Riggi}, Simone and {Ross}, Kathryn and {Rowell}, Gavin and {Rudnick}, Lawrence and {Sano}, Hidetoshi and {Sasaki}, Manami and {Soria}, Roberto and {Urosevic}, Dejan and {Vukotic}, Branislav and {West}, Jennifer L.},
        title = "{Teleios (G305.4-2.2) -- the mystery of a perfectly shaped new Galactic supernova remnant}",
      journal = {},
         year = 2025,
        month = may,
          eid = {arXiv:2505.04041},
        pages = {arXiv:2505.04041},
          doi = {10.48550/arXiv.2505.04041},
archivePrefix = {arXiv},
       eprint = {2505.04041},
 primaryClass = {astro-ph.HE},
       adsurl = {https://ui.adsabs.harvard.edu/abs/2025arXiv250504041F}
}

@book{book1,
	author = {Filipovi{\'c}, Miroslav D and Tothill, Nicholas F H},
	doi = {10.1088/2514-3433/ac087e},
	isbn = {978-0-7503-2340-6},
	publisher = {IOP Publishing},
	series = {2514-3433},
	title = {Principles of Multimessenger Astronomy},
	url = {https://dx.doi.org/10.1088/2514-3433/ac087e},
	year = {2021}}

@book{book2,
	doi = {10.1088/2514-3433/ac2256},
	editor = {Filipovi{\'c}, Miroslav D and Tothill, Nicholas F H},
	isbn = {978-0-7503-2344-4},
	publisher = {IOP Publishing},
	series = {2514-3433},
	title = {Multimessenger Astronomy in Practice},
	url = {https://dx.doi.org/10.1088/2514-3433/ac2256},
	year = {2021}}

@ARTICLE{Cotton_2024MNRAS.529.2443C,
       author = {{Cotton}, W.~D. and {Filipovi{\'c}}, M.~D. and {Camilo}, F. and {Indebetouw}, R. and {Alsaberi}, R.~Z.~E. and {Anih}, J.~O. and {Baker}, M. and {Bastian}, T.~S. and {Boji{\v{c}}i{\'c}}, I. and {Carli}, E. and {Cavallaro}, F. and {Crawford}, E.~J. and {Dai}, S. and {Haberl}, F. and {Levin}, L. and {Luken}, K. and {Pennock}, C.~M. and {Rajabpour}, N. and {Stappers}, B.~W. and {van Loon}, J. Th and {Zijlstra}, A.~A. and {Buchner}, S. and {Geyer}, M. and {Goedhart}, S. and {Serylak}, M.},
        title = "{The MeerKAT 1.3 GHz Survey of the Small Magellanic Cloud}",
      journal = {MNRAS},
         year = 2024,
        month = apr,
       volume = {529},
       number = {3},
        pages = {2443-2472},
          doi = {10.1093/mnras/stae277},
archivePrefix = {arXiv},
       eprint = {2401.11024},
 primaryClass = {astro-ph.GA},
       adsurl = {https://ui.adsabs.harvard.edu/abs/2024MNRAS.529.2443C}
}

@ARTICLE{Acero_2016ApJS..224....8A,
       author = {{Acero}, F. and {Ackermann}, M. and {Ajello}, M. and {Baldini}, L. and {Ballet}, J. and {Barbiellini}, G. and {Bastieri}, D. and {Bellazzini}, R. and {Bissaldi}, E. and {Blandford}, R.~D. and {Bloom}, E.~D. and {Bonino}, R. and {Bottacini}, E. and {Brandt}, T.~J. and {Bregeon}, J. and {Bruel}, P. and {Buehler}, R. and {Buson}, S. and {Caliandro}, G.~A. and {Cameron}, R.~A. and {Caputo}, R. and {Caragiulo}, M. and {Caraveo}, P.~A. and {Casandjian}, J.~M. and {Cavazzuti}, E. and {Cecchi}, C. and {Chekhtman}, A. and {Chiang}, J. and {Chiaro}, G. and {Ciprini}, S. and {Claus}, R. and {Cohen}, J.~M. and {Cohen-Tanugi}, J. and {Cominsky}, L.~R. and {Condon}, B. and {Conrad}, J. and {Cutini}, S. and {D'Ammando}, F. and {de Angelis}, A. and {de Palma}, F. and {Desiante}, R. and {Digel}, S.~W. and {Di Venere}, L. and {Drell}, P.~S. and {Drlica-Wagner}, A. and {Favuzzi}, C. and {Ferrara}, E.~C. and {Franckowiak}, A. and {Fukazawa}, Y. and {Funk}, S. and {Fusco}, P. and {Gargano}, F. and {Gasparrini}, D. and {Giglietto}, N. and {Giommi}, P. and {Giordano}, F. and {Giroletti}, M. and {Glanzman}, T. and {Godfrey}, G. and {Gomez-Vargas}, G.~A. and {Grenier}, I.~A. and {Grondin}, M. -H. and {Guillemot}, L. and {Guiriec}, S. and {Gustafsson}, M. and {Hadasch}, D. and {Harding}, A.~K. and {Hayashida}, M. and {Hays}, E. and {Hewitt}, J.~W. and {Hill}, A.~B. and {Horan}, D. and {Hou}, X. and {Iafrate}, G. and {Jogler}, T. and {J{\'o}hannesson}, G. and {Johnson}, A.~S. and {Kamae}, T. and {Katagiri}, H. and {Kataoka}, J. and {Katsuta}, J. and {Kerr}, M. and {Kn{\"o}dlseder}, J. and {Kocevski}, D. and {Kuss}, M. and {Laffon}, H. and {Lande}, J. and {Larsson}, S. and {Latronico}, L. and {Lemoine-Goumard}, M. and {Li}, J. and {Li}, L. and {Longo}, F. and {Loparco}, F. and {Lovellette}, M.~N. and {Lubrano}, P. and {Magill}, J. and {Maldera}, S. and {Marelli}, M. and {Mayer}, M. and {Mazziotta}, M.~N. and {Michelson}, P.~F. and {Mitthumsiri}, W. and {Mizuno}, T. and {Moiseev}, A.~A. and {Monzani}, M.~E. and {Moretti}, E. and {Morselli}, A. and {Moskalenko}, I.~V. and {Murgia}, S. and {Nemmen}, R. and {Nuss}, E. and {Ohsugi}, T. and {Omodei}, N. and {Orienti}, M. and {Orlando}, E. and {Ormes}, J.~F. and {Paneque}, D. and {Perkins}, J.~S. and {Pesce-Rollins}, M. and {Petrosian}, V. and {Piron}, F. and {Pivato}, G. and {Porter}, T.~A. and {Rain{\`o}}, S. and {Rando}, R. and {Razzano}, M. and {Razzaque}, S. and {Reimer}, A. and {Reimer}, O. and {Renaud}, M. and {Reposeur}, T. and {Rousseau}, R. and {Saz Parkinson}, P.~M. and {Schmid}, J. and {Schulz}, A. and {Sgr{\`o}}, C. and {Siskind}, E.~J. and {Spada}, F. and {Spandre}, G. and {Spinelli}, P. and {Strong}, A.~W. and {Suson}, D.~J. and {Tajima}, H. and {Takahashi}, H. and {Tanaka}, T. and {Thayer}, J.~B. and {Thompson}, D.~J. and {Tibaldo}, L. and {Tibolla}, O. and {Torres}, D.~F. and {Tosti}, G. and {Troja}, E. and {Uchiyama}, Y. and {Vianello}, G. and {Wells}, B. and {Wood}, K.~S. and {Wood}, M. and {Yassine}, M. and {den Hartog}, P.~R. and {Zimmer}, S.},
        title = "{The First Fermi LAT Supernova Remnant Catalog}",
      journal = {\apjs},
         year = 2016,
        month = may,
       volume = {224},
       number = {1},
          eid = {8},
        pages = {8},
          doi = {10.3847/0067-0049/224/1/8},
archivePrefix = {arXiv},
       eprint = {1511.06778},
 primaryClass = {astro-ph.HE},
       adsurl = {https://ui.adsabs.harvard.edu/abs/2016ApJS..224....8A}
}

@ARTICLE{Abdollahi_2020ApJS..247...33A,
       author = {{Abdollahi}, S. and {Acero}, F. and {Ackermann}, M. and {Ajello}, M. and {Atwood}, W.~B. and {Axelsson}, M. and {Baldini}, L. and {Ballet}, J. and {Barbiellini}, G. and {Bastieri}, D. and {Becerra Gonzalez}, J. and {Bellazzini}, R. and {Berretta}, A. and {Bissaldi}, E. and {Blandford}, R.~D. and {Bloom}, E.~D. and {Bonino}, R. and {Bottacini}, E. and {Brandt}, T.~J. and {Bregeon}, J. and {Bruel}, P. and {Buehler}, R. and {Burnett}, T.~H. and {Buson}, S. and {Cameron}, R.~A. and {Caputo}, R. and {Caraveo}, P.~A. and {Casandjian}, J.~M. and {Castro}, D. and {Cavazzuti}, E. and {Charles}, E. and {Chaty}, S. and {Chen}, S. and {Cheung}, C.~C. and {Chiaro}, G. and {Ciprini}, S. and {Cohen-Tanugi}, J. and {Cominsky}, L.~R. and {Coronado-Bl{\'a}zquez}, J. and {Costantin}, D. and {Cuoco}, A. and {Cutini}, S. and {D'Ammando}, F. and {DeKlotz}, M. and {de la Torre Luque}, P. and {de Palma}, F. and {Desai}, A. and {Digel}, S.~W. and {Di Lalla}, N. and {Di Mauro}, M. and {Di Venere}, L. and {Dom{\'\i}nguez}, A. and {Dumora}, D. and {Fana Dirirsa}, F. and {Fegan}, S.~J. and {Ferrara}, E.~C. and {Franckowiak}, A. and {Fukazawa}, Y. and {Funk}, S. and {Fusco}, P. and {Gargano}, F. and {Gasparrini}, D. and {Giglietto}, N. and {Giommi}, P. and {Giordano}, F. and {Giroletti}, M. and {Glanzman}, T. and {Green}, D. and {Grenier}, I.~A. and {Griffin}, S. and {Grondin}, M. -H. and {Grove}, J.~E. and {Guiriec}, S. and {Harding}, A.~K. and {Hayashi}, K. and {Hays}, E. and {Hewitt}, J.~W. and {Horan}, D. and {J{\'o}hannesson}, G. and {Johnson}, T.~J. and {Kamae}, T. and {Kerr}, M. and {Kocevski}, D. and {Kovac'evic'}, M. and {Kuss}, M. and {Landriu}, D. and {Larsson}, S. and {Latronico}, L. and {Lemoine-Goumard}, M. and {Li}, J. and {Liodakis}, I. and {Longo}, F. and {Loparco}, F. and {Lott}, B. and {Lovellette}, M.~N. and {Lubrano}, P. and {Madejski}, G.~M. and {Maldera}, S. and {Malyshev}, D. and {Manfreda}, A. and {Marchesini}, E.~J. and {Marcotulli}, L. and {Mart{\'\i}-Devesa}, G. and {Martin}, P. and {Massaro}, F. and {Mazziotta}, M.~N. and {McEnery}, J.~E. and {Mereu}, I. and {Meyer}, M. and {Michelson}, P.~F. and {Mirabal}, N. and {Mizuno}, T. and {Monzani}, M.~E. and {Morselli}, A. and {Moskalenko}, I.~V. and {Negro}, M. and {Nuss}, E. and {Ojha}, R. and {Omodei}, N. and {Orienti}, M. and {Orlando}, E. and {Ormes}, J.~F. and {Palatiello}, M. and {Paliya}, V.~S. and {Paneque}, D. and {Pei}, Z. and {Pe{\~n}a-Herazo}, H. and {Perkins}, J.~S. and {Persic}, M. and {Pesce-Rollins}, M. and {Petrosian}, V. and {Petrov}, L. and {Piron}, F. and {Poon}, H. and {Porter}, T.~A. and {Principe}, G. and {Rain{\`o}}, S. and {Rando}, R. and {Razzano}, M. and {Razzaque}, S. and {Reimer}, A. and {Reimer}, O. and {Remy}, Q. and {Reposeur}, T. and {Romani}, R.~W. and {Saz Parkinson}, P.~M. and {Schinzel}, F.~K. and {Serini}, D. and {Sgr{\`o}}, C. and {Siskind}, E.~J. and {Smith}, D.~A. and {Spandre}, G. and {Spinelli}, P. and {Strong}, A.~W. and {Suson}, D.~J. and {Tajima}, H. and {Takahashi}, M.~N. and {Tak}, D. and {Thayer}, J.~B. and {Thompson}, D.~J. and {Tibaldo}, L. and {Torres}, D.~F. and {Torresi}, E. and {Valverde}, J. and {Van Klaveren}, B. and {van Zyl}, P. and {Wood}, K. and {Yassine}, M. and {Zaharijas}, G.},
        title = "{Fermi Large Area Telescope Fourth Source Catalog}",
      journal = {\apjs},
         year = 2020,
        month = mar,
       volume = {247},
       number = {1},
          eid = {33},
        pages = {33},
          doi = {10.3847/1538-4365/ab6bcb},
archivePrefix = {arXiv},
       eprint = {1902.10045},
 primaryClass = {astro-ph.HE},
       adsurl = {https://ui.adsabs.harvard.edu/abs/2020ApJS..247...33A}
}

@misc{Ballet_2024_Fermi-4FGL-DR4,
      title={Fermi Large Area Telescope Fourth Source Catalog Data Release 4 (4FGL-DR4)}, 
      author={J. Ballet and P. Bruel and T. H. Burnett and B. Lott and The Fermi-LAT collaboration},
      year={2024},
      eprint={2307.12546},
      archivePrefix={arXiv},
      primaryClass={astro-ph.HE},
      url={https://arxiv.org/abs/2307.12546}, 
}

@INPROCEEDINGS{Fermipy_2017,
       author = {{Wood}, M. and {Caputo}, R. and {Charles}, E. and {Di Mauro}, M. and {Magill}, J. and {Perkins}, J.~S. and {Fermi-LAT Collaboration}},
        title = "{Fermipy: An open-source Python package for analysis of Fermi-LAT Data}",
    booktitle = {35th International Cosmic Ray Conference (ICRC2017)},
         year = 2017,
       series = {International Cosmic Ray Conference},
       volume = {301},
        month = jul,
          eid = {824},
        pages = {824},
          doi = {10.22323/1.301.0824},
archivePrefix = {arXiv},
       eprint = {1707.09551},
 primaryClass = {astro-ph.IM},
       adsurl = {https://ui.adsabs.harvard.edu/abs/2017ICRC...35..824W}
}

@misc{Atwood_2013,
      title={Pass 8: Toward the Full Realization of the Fermi-LAT Scientific Potential}, 
      author={W. Atwood and A. Albert and L. Baldini and M. Tinivella and J. Bregeon and M. Pesce-Rollins and C. Sgrò and P. Bruel and E. Charles and A. Drlica-Wagner and A. Franckowiak and T. Jogler and L. Rochester and T. Usher and M. Wood and J. Cohen-Tanugi and S. Zimmer},
      year={2013},
      eprint={1303.3514},
      archivePrefix={arXiv},
      primaryClass={astro-ph.IM}
}

@ARTICLE{HESS_2018A&A...612A...3H,
       author = {{H.~E.~S.~S. Collaboration} and {Abdalla}, H. and {Abramowski}, A. and {Aharonian}, F. and {Ait Benkhali}, F. and {Ang{\"u}ner}, E.~O. and {Arakawa}, M. and {Arrieta}, M. and {Aubert}, P. and {Backes}, M. and {Balzer}, A. and {Barnard}, M. and {Becherini}, Y. and {Becker Tjus}, J. and {Berge}, D. and {Bernhard}, S. and {Bernl{\"o}hr}, K. and {Blackwell}, R. and {B{\"o}ttcher}, M. and {Boisson}, C. and {Bolmont}, J. and {Bonnefoy}, S. and {Bordas}, P. and {Bregeon}, J. and {Brun}, F. and {Brun}, P. and {Bryan}, M. and {B{\"u}chele}, M. and {Bulik}, T. and {Capasso}, M. and {Caroff}, S. and {Carosi}, A. and {Casanova}, S. and {Cerruti}, M. and {Chakraborty}, N. and {Chaves}, R.~C.~G. and {Chen}, A. and {Chevalier}, J. and {Colafrancesco}, S. and {Condon}, B. and {Conrad}, J. and {Davids}, I.~D. and {Decock}, J. and {Deil}, C. and {Devin}, J. and {deWilt}, P. and {Dirson}, L. and {Djannati-Ata{\"\i}}, A. and {Donath}, A. and {Drury}, L.~O. 'C. and {Dutson}, K. and {Dyks}, J. and {Edwards}, T. and {Egberts}, K. and {Emery}, G. and {Ernenwein}, J. -P. and {Eschbach}, S. and {Farnier}, C. and {Fegan}, S. and {Fernandes}, M.~V. and {Fernandez}, D. and {Fiasson}, A. and {Fontaine}, G. and {Funk}, S. and {F{\"u}{\ss}ling}, M. and {Gabici}, S. and {Gallant}, Y.~A. and {Garrigoux}, T. and {Gat{\'e}}, F. and {Giavitto}, G. and {Giebels}, B. and {Glawion}, D. and {Glicenstein}, J.~F. and {Gottschall}, D. and {Grondin}, M. -H. and {Hahn}, J. and {Haupt}, M. and {Hawkes}, J. and {Heinzelmann}, G. and {Henri}, G. and {Hermann}, G. and {Hinton}, J.~A. and {Hofmann}, W. and {Hoischen}, C. and {Holch}, T.~L. and {Holler}, M. and {Horns}, D. and {Ivascenko}, A. and {Iwasaki}, H. and {Jacholkowska}, A. and {Jamrozy}, M. and {Jankowsky}, D. and {Jankowsky}, F. and {Jingo}, M. and {Jouvin}, L. and {Jung-Richardt}, I. and {Kastendieck}, M.~A. and {Katarzy{\'n}ski}, K. and {Katsuragawa}, M. and {Katz}, U. and {Kerszberg}, D. and {Khangulyan}, D. and {Kh{\'e}lifi}, B. and {King}, J. and {Klepser}, S. and {Klochkov}, D. and {Klu{\'z}niak}, W. and {Komin}, Nu. and {Kosack}, K. and {Krakau}, S. and {Kraus}, M. and {Kr{\"u}ger}, P.~P. and {Laffon}, H. and {Lamanna}, G. and {Lau}, J. and {Lees}, J. -P. and {Lefaucheur}, J. and {Lemi{\`e}re}, A. and {Lemoine-Goumard}, M. and {Lenain}, J. -P. and {Leser}, E. and {Lohse}, T. and {Lorentz}, M. and {Liu}, R. and {L{\'o}pez-Coto}, R. and {Lypova}, I. and {Malyshev}, D. and {Marandon}, V. and {Marcowith}, A. and {Mariaud}, C. and {Marx}, R. and {Maurin}, G. and {Maxted}, N. and {Mayer}, M. and {Meintjes}, P.~J. and {Meyer}, M. and {Mitchell}, A.~M.~W. and {Moderski}, R. and {Mohamed}, M. and {Mohrmann}, L. and {Mor{\r{a}}}, K. and {Moulin}, E. and {Murach}, T. and {Nakashima}, S. and {de Naurois}, M. and {Ndiyavala}, H. and {Niederwanger}, F. and {Niemiec}, J. and {Oakes}, L. and {O'Brien}, P. and {Odaka}, H. and {Ohm}, S. and {Ostrowski}, M. and {Oya}, I. and {Padovani}, M. and {Panter}, M. and {Parsons}, R.~D. and {Pekeur}, N.~W. and {Pelletier}, G. and {Perennes}, C. and {Petrucci}, P. -O. and {Peyaud}, B. and {Piel}, Q. and {Pita}, S. and {Poireau}, V. and {Poon}, H. and {Prokhorov}, D. and {Prokoph}, H. and {P{\"u}hlhofer}, G. and {Punch}, M. and {Quirrenbach}, A. and {Raab}, S. and {Rauth}, R. and {Reimer}, A. and {Reimer}, O. and {Renaud}, M. and {de los Reyes}, R. and {Rieger}, F. and {Rinchiuso}, L. and {Romoli}, C. and {Rowell}, G. and {Rudak}, B. and {Rulten}, C.~B. and {Safi-Harb}, S. and {Sahakian}, V. and {Saito}, S. and {Sanchez}, D.~A. and {Santangelo}, A. and {Sasaki}, M. and {Schlickeiser}, R. and {Sch{\"u}ssler}, F. and {Schulz}, A. and {Schwanke}, U. and {Schwemmer}, S. and {Seglar-Arroyo}, M. and {Settimo}, M. and {Seyffert}, A.~S. and {Shafi}, N. and {Shilon}, I. and {Shiningayamwe}, K.},
        title = "{Population study of Galactic supernova remnants at very high {\ensuremath{\gamma}}-ray energies with H.E.S.S.}",
      journal = {\aap},
         year = 2018,
        month = apr,
       volume = {612},
          eid = {A3},
        pages = {A3},
          doi = {10.1051/0004-6361/201732125},
archivePrefix = {arXiv},
       eprint = {1802.05172},
 primaryClass = {astro-ph.HE},
       adsurl = {https://ui.adsabs.harvard.edu/abs/2018A&A...612A...3H}
}

@ARTICLE{Ball_2023,
       author = {{Ball}, Brianna D. and {Kothes}, Roland and {Rosolowsky}, Erik and {West}, Jennifer and {Becker}, Werner and {Filipovi{\'c}}, Miroslav D. and {Gaensler}, B.~M. and {Hopkins}, Andrew M. and {Koribalski}, B{\"a}rbel and {Landecker}, Tom and {Leahy}, Denis and {Marvil}, Joshua and {Sun}, Xiaohui and {Bufano}, Filomena and {Carretti}, Ettore and {Ingallinera}, Adriano and {Van Eck}, Cameron L. and {Willis}, Tony},
        title = "{A catalogue of radio supernova remnants and candidate supernova remnants in the EMU/POSSUM Galactic pilot field}",
      journal = {\mnras},
         year = 2023,
        month = sep,
       volume = {524},
       number = {1},
        pages = {1396-1421},
          doi = {10.1093/mnras/stad1953},
archivePrefix = {arXiv},
       eprint = {2307.01948},
 primaryClass = {astro-ph.GA},
       adsurl = {https://ui.adsabs.harvard.edu/abs/2023MNRAS.524.1396B}
}

@ARTICLE{Ball_2025ApJ...988...75B,
       author = {{Ball}, B.~D. and {Kothes}, R. and {Rosolowsky}, E. and {Burger-Scheidlin}, C. and {Filipovi{\'c}}, M.~D. and {Lazarevi{\'c}}, S. and {Smeaton}, Z.~J. and {Becker}, W. and {Carretti}, E. and {Gaensler}, B.~M. and {Hopkins}, A.~M. and {Leahy}, D. and {Tahani}, M. and {West}, J.~L. and {Anderson}, C.~S. and {Loru}, S. and {Ma}, Y.~K. and {McClure-Griffiths}, N.~M. and {Micha{\l}owski}, M.~J.},
        title = "{A Catalog of Galactic Supernova Remnants and Supernova Remnant Candidates from the EMU/POSSUM Radio Sky Surveys. I.}",
      journal = {\apj},
         year = 2025,
        month = jul,
       volume = {988},
       number = {1},
          eid = {75},
        pages = {75},
          doi = {10.3847/1538-4357/addc63},
       adsurl = {https://ui.adsabs.harvard.edu/abs/2025ApJ...988...75B}
}

@ARTICLE{GreenA_2014,
       author = {{Green}, A.~J. and {Reeves}, S.~N. and {Murphy}, T.},
        title = "{The Second Epoch Molonglo Galactic Plane Survey: Images and Candidate Supernova Remnants}",
      journal = {\pasa},
         year = 2014,
        month = nov,
       volume = {31},
          eid = {e042},
        pages = {e042},
          doi = {10.1017/pasa.2014.37},
archivePrefix = {arXiv},
       eprint = {1410.8247},
 primaryClass = {astro-ph.GA},
       adsurl = {https://ui.adsabs.harvard.edu/abs/2014PASA...31...42G}
}

@ARTICLE{Green_2017,
       author = {{Green}, D.~A.},
        title = "{VizieR Online Data Catalog: A Catalogue of Galactic Supernova Remnants (Green 2017)}",
      journal = {VizieR Online Data Catalog},
         year = 2017,
        month = jun,
          eid = {VII/278},
        pages = {VII/278},
       adsurl = {https://ui.adsabs.harvard.edu/abs/2017yCat.7278....0G}
}

@ARTICLE{Green_2025JApA...46...14G,
       author = {{Green}, D.~A.},
        title = "{An updated catalogue of 310 Galactic supernova remnants and their statistical properties}",
      journal = {Journal of Astrophysics and Astronomy},
         year = 2025,
        month = jan,
       volume = {46},
       number = {1},
          eid = {14},
        pages = {14},
          doi = {10.1007/s12036-024-10038-4},
archivePrefix = {arXiv},
       eprint = {2411.03367},
 primaryClass = {astro-ph.GA},
       adsurl = {https://ui.adsabs.harvard.edu/abs/2025JApA...46...14G}
}

@ARTICLE{Ferrand_2012,
       author = {{Ferrand}, Gilles and {Safi-Harb}, Samar},
        title = "{A census of high-energy observations of Galactic supernova remnants}",
      journal = {Advances in Space Research},
         year = 2012,
        month = may,
       volume = {49},
       number = {9},
        pages = {1313-1319},
          doi = {10.1016/j.asr.2012.02.004},
archivePrefix = {arXiv},
       eprint = {1202.0245},
 primaryClass = {astro-ph.HE},
       adsurl = {https://ui.adsabs.harvard.edu/abs/2012AdSpR..49.1313F}
}

@misc{EMU_2022,
    note = "Data collection",
    author = {Hopkins, Andrew and Norris, Ray and Vernstrom, Tessa and Kapinska, Anna and Marvil, Josh},
    title = "{ASKAP Data Products for Project AS201 (EMU): images and visibilities v1}",
    publisher = {CSIRO}, 
    year = 2022,
    adsurl = {http://hdl.handle.net/102.100.100/479788?index=1}
}

@ARTICLE{Devin_2020,
       author = {{Devin}, J. and {Lemoine-Goumard}, M. and {Grondin}, M. -H. and {Castro}, D. and {Ballet}, J. and {Cohen}, J. and {Hewitt}, J.~W.},
        title = "{High-energy gamma-ray study of the dynamically young SNR G150.3+4.5}",
      journal = {\aap},
         year = 2020,
        month = nov,
       volume = {643},
          eid = {A28},
        pages = {A28},
          doi = {10.1051/0004-6361/202038503},
archivePrefix = {arXiv},
       eprint = {2009.08397},
 primaryClass = {astro-ph.HE},
       adsurl = {https://ui.adsabs.harvard.edu/abs/2020A&A...643A..28D}
}

@ARTICLE{Astropy_2022,
        author = {{Astropy Collaboration} and {Price-Whelan}, Adrian M. and {Lim}, Pey Lian and {Earl}, Nicholas and {Starkman}, Nathaniel and {Bradley}, Larry and {Shupe}, David L. and {Patil}, Aarya A. and {Corrales}, Lia and {Brasseur}, C.~E. and {N{\"o}the}, Maximilian and {Donath}, Axel and {Tollerud}, Erik and {Morris}, Brett M. and {Ginsburg}, Adam and {Vaher}, Eero and {Weaver}, Benjamin A. and {Tocknell}, James and {Jamieson}, William and {van Kerkwijk}, Marten H. and {Robitaille}, Thomas P. and {Merry}, Bruce and {Bachetti}, Matteo and {G{\"u}nther}, H. Moritz and {Aldcroft}, Thomas L. and {Alvarado-Montes}, Jaime A. and {Archibald}, Anne M. and {B{\'o}di}, Attila and {Bapat}, Shreyas and {Barentsen}, Geert and {Baz{\'a}n}, Juanjo and {Biswas}, Manish and {Boquien}, M{\'e}d{\'e}ric and {Burke}, D.~J. and {Cara}, Daria and {Cara}, Mihai and {Conroy}, Kyle E. and {Conseil}, Simon and {Craig}, Matthew W. and {Cross}, Robert M. and {Cruz}, Kelle L. and {D'Eugenio}, Francesco and {Dencheva}, Nadia and {Devillepoix}, Hadrien A.~R. and {Dietrich}, J{\"o}rg P. and {Eigenbrot}, Arthur Davis and {Erben}, Thomas and {Ferreira}, Leonardo and {Foreman-Mackey}, Daniel and {Fox}, Ryan and {Freij}, Nabil and {Garg}, Suyog and {Geda}, Robel and {Glattly}, Lauren and {Gondhalekar}, Yash and {Gordon}, Karl D. and {Grant}, David and {Greenfield}, Perry and {Groener}, Austen M. and {Guest}, Steve and {Gurovich}, Sebastian and {Handberg}, Rasmus and {Hart}, Akeem and {Hatfield-Dodds}, Zac and {Homeier}, Derek and {Hosseinzadeh}, Griffin and {Jenness}, Tim and {Jones}, Craig K. and {Joseph}, Prajwel and {Kalmbach}, J. Bryce and {Karamehmetoglu}, Emir and {Ka{\l}uszy{\'n}ski}, Miko{\l}aj and {Kelley}, Michael S.~P. and {Kern}, Nicholas and {Kerzendorf}, Wolfgang E. and {Koch}, Eric W. and {Kulumani}, Shankar and {Lee}, Antony and {Ly}, Chun and {Ma}, Zhiyuan and {MacBride}, Conor and {Maljaars}, Jakob M. and {Muna}, Demitri and {Murphy}, N.~A. and {Norman}, Henrik and {O'Steen}, Richard and {Oman}, Kyle A. and {Pacifici}, Camilla and {Pascual}, Sergio and {Pascual-Granado}, J. and {Patil}, Rohit R. and {Perren}, Gabriel I. and {Pickering}, Timothy E. and {Rastogi}, Tanuj and {Roulston}, Benjamin R. and {Ryan}, Daniel F. and {Rykoff}, Eli S. and {Sabater}, Jose and {Sakurikar}, Parikshit and {Salgado}, Jes{\'u}s and {Sanghi}, Aniket and {Saunders}, Nicholas and {Savchenko}, Volodymyr and {Schwardt}, Ludwig and {Seifert-Eckert}, Michael and {Shih}, Albert Y. and {Jain}, Anany Shrey and {Shukla}, Gyanendra and {Sick}, Jonathan and {Simpson}, Chris and {Singanamalla}, Sudheesh and {Singer}, Leo P. and {Singhal}, Jaladh and {Sinha}, Manodeep and {Sip{\H{o}}cz}, Brigitta M. and {Spitler}, Lee R. and {Stansby}, David and {Streicher}, Ole and {{\v{S}}umak}, Jani and {Swinbank}, John D. and {Taranu}, Dan S. and {Tewary}, Nikita and {Tremblay}, Grant R. and {de Val-Borro}, Miguel and {Van Kooten}, Samuel J. and {Vasovi{\'c}}, Zlatan and {Verma}, Shresth and {de Miranda Cardoso}, Jos{\'e} Vin{\'\i}cius and {Williams}, Peter K.~G. and {Wilson}, Tom J. and {Winkel}, Benjamin and {Wood-Vasey}, W.~M. and {Xue}, Rui and {Yoachim}, Peter and {Zhang}, Chen and {Zonca}, Andrea and {Astropy Project Contributors}},
        title = "{The Astropy Project: Sustaining and Growing a Community-oriented Open-source Project and the Latest Major Release (v5.0) of the Core Package}",
        journal = {\apj},
        year = 2022,
        month = aug,
        volume = {935},
        number = {2},
        eid = {167},
        pages = {167},
        doi = {10.3847/1538-4357/ac7c74},
        archivePrefix = {arXiv},
        eprint = {2206.14220},
        primaryClass = {astro-ph.IM},
        adsurl = {https://ui.adsabs.harvard.edu/abs/2022ApJ...935..167A}
}

@Article{Numpy_2020,
     title         = {Array programming with {NumPy}},
     author        = {Charles R. Harris and K. Jarrod Millman and St{\'{e}}fan J.
                     van der Walt and Ralf Gommers and Pauli Virtanen and David
                     Cournapeau and Eric Wieser and Julian Taylor and Sebastian
                     Berg and Nathaniel J. Smith and Robert Kern and Matti Picus
                     and Stephan Hoyer and Marten H. van Kerkwijk and Matthew
                     Brett and Allan Haldane and Jaime Fern{\'{a}}ndez del
                     R{\'{i}}o and Mark Wiebe and Pearu Peterson and Pierre
                     G{\'{e}}rard-Marchant and Kevin Sheppard and Tyler Reddy and
                     Warren Weckesser and Hameer Abbasi and Christoph Gohlke and
                     Travis E. Oliphant},
     year          = {2020},
     month         = sep,
     journal       = {Nature},
     volume        = {585},
     number        = {7825},
     pages         = {357--362},
     doi           = {10.1038/s41586-020-2649-2},
     publisher     = {Springer Science and Business Media {LLC}},
     url           = {https://doi.org/10.1038/s41586-020-2649-2}
    }

@article{Arias_2022,
	doi = {10.1051/0004-6361/202244369},
	url = {https://doi.org/10.1051%2F0004-6361%2F202244369},
	year = 2022,
	month = {nov},
	publisher = {{EDP} Sciences},
	volume = {667},
	pages = {A71},
	author = {M. Arias and A. Botteon and C. G. Bassa and S. van der Jagt and R. J. van Weeren and S. P. O'Sullivan and Q. Bosschaart and R. S. Dullaart and M. J. Hardcastle and J. W. T. Hessels and T. Shimwell and M. M. Slob and J. A. Sturm and C. Tasse and N. C. M. A. Theijssen and J. Vink},
  
	title = {Possible discovery of Calvera's supernova remnant}, 
	journal = {\aap}
}

@article{Araya_2023,
       author = {{Araya}, M.},
        title = "{Fermi-LAT detection of G118.4+37.0: a supernova remnant in the Galactic halo seen around the Calvera pulsar}",
      journal = {\mnras},
         year = 2023,
        month = jan,
       volume = {518},
       number = {3},
        pages = {4132-4137},
          doi = {10.1093/mnras/stac3337},
archivePrefix = {arXiv},
       eprint = {2210.16340},
 primaryClass = {astro-ph.HE},
       adsurl = {https://ui.adsabs.harvard.edu/abs/2023MNRAS.518.4132A}
}

@ARTICLE{2007MNRAS.375...92R,
       author = {{Reynoso}, E.~M. and {Green}, A.~J.},
        title = "{G332.5-5.6, a new Galactic supernova remnant}",
      journal = {\mnras},
         year = 2007,
        month = feb,
       volume = {375},
       number = {1},
        pages = {92-98},
          doi = {10.1111/j.1365-2966.2006.11264.x},
archivePrefix = {arXiv},
       eprint = {astro-ph/0611234},
 primaryClass = {astro-ph},
       adsurl = {https://ui.adsabs.harvard.edu/abs/2007MNRAS.375...92R}
}

@ARTICLE{1994MNRAS.270..106M,
       author = {{Milne}, D.~K. and {Haynes}, R.~F.},
        title = "{PKS 1209-51/52 : a supernova remnant with a well-defined magnetic field.}",
      journal = {\mnras},
         year = 1994,
        month = sep,
       volume = {270},
        pages = {106-114},
          doi = {10.1093/mnras/270.1.106},
       adsurl = {https://ui.adsabs.harvard.edu/abs/1994MNRAS.270..106M}
}

@ARTICLE{DeLaney_2002ApJ...580..914D,
       author = {{DeLaney}, Tracey and {Koralesky}, Barron and {Rudnick}, Lawrence and {Dickel}, John R.},
        title = "{Radio Spectral Index Variations and Physical Conditions in Kepler's Supernova Remnant}",
      journal = {\apj},
         year = 2002,
        month = dec,
       volume = {580},
       number = {2},
        pages = {914-927},
          doi = {10.1086/343787},
archivePrefix = {arXiv},
       eprint = {astro-ph/0210355},
 primaryClass = {astro-ph},
       adsurl = {https://ui.adsabs.harvard.edu/abs/2002ApJ...580..914D}
}

@ARTICLE{2006_Kothes,
       author = {{Kothes}, R. and {Fedotov}, K. and {Foster}, T.~J. and {Uyan{\i}ker}, B.},
        title = "{A catalogue of Galactic supernova remnants from the Canadian Galactic plane survey. I. Flux densities, spectra, and polarization characteristics}",
      journal = {\aap},
         year = 2006,
        month = oct,
       volume = {457},
       number = {3},
        pages = {1081-1093},
          doi = {10.1051/0004-6361:20065062},
       adsurl = {https://ui.adsabs.harvard.edu/abs/2006A&A...457.1081K}
}

@ARTICLE{Gao_2011AA...529A.159G,
       author = {{Gao}, X.~Y. and {Han}, J.~L. and {Reich}, W. and {Reich}, P. and {Sun}, X.~H. and {Xiao}, L.},
        title = "{A Sino-German {\ensuremath{\lambda}}6 cm polarization survey of the Galactic plane. V. Large supernova remnants}",
      journal = {\aap},
         year = 2011,
        month = may,
       volume = {529},
          eid = {A159},
        pages = {A159},
          doi = {10.1051/0004-6361/201016311},
archivePrefix = {arXiv},
       eprint = {1102.4503},
 primaryClass = {astro-ph.GA},
       adsurl = {https://ui.adsabs.harvard.edu/abs/2011A&A...529A.159G}
}

@ARTICLE{2004AA...426..909U,
       author = {{Uyan{\i}ker}, B. and {Reich}, W. and {Yar}, A. and {F{\"u}rst}, E.},
        title = "{Radio emission from the Cygnus Loop and its spectral characteristics}",
      journal = {\aap},
         year = 2004,
        month = nov,
       volume = {426},
        pages = {909-924},
          doi = {10.1051/0004-6361:200400040},
archivePrefix = {arXiv},
       eprint = {astro-ph/0409176},
 primaryClass = {astro-ph},
       adsurl = {https://ui.adsabs.harvard.edu/abs/2004A&A...426..909U}
}

@ARTICLE{2006AA...447..937S,
       author = {{Sun}, X.~H. and {Reich}, W. and {Han}, J.~L. and {Reich}, P. and {Wielebinski}, R.},
        title = "{New {\ensuremath{\lambda}}6 cm observations of the Cygnus Loop}",
      journal = {\aap},
         year = 2006,
        month = mar,
       volume = {447},
       number = {3},
        pages = {937-947},
          doi = {10.1051/0004-6361:20054133},
archivePrefix = {arXiv},
       eprint = {astro-ph/0510509},
 primaryClass = {astro-ph},
       adsurl = {https://ui.adsabs.harvard.edu/abs/2006A&A...447..937S}
}

@article{Araya_2022,
       author = {{Araya}, M. and {Hurley-Walker}, N. and {Quir{\'o}s-Araya}, S.},
        title = "{G17.8 + 16.7: A new supernova remnant}",
      journal = {\mnras},
         year = 2022,
        month = feb,
       volume = {510},
       number = {2},
        pages = {2920-2927},
          doi = {10.1093/mnras/stab3550},
archivePrefix = {arXiv},
       eprint = {2112.01985},
 primaryClass = {astro-ph.HE},
       adsurl = {https://ui.adsabs.harvard.edu/abs/2022MNRAS.510.2920A}
}

@ARTICLE{Fesen_2024,
       author = {{Fesen}, Robert A. and {Drechsler}, Marcel and {Strottner}, Xavier and {Falls}, Bray and {Sainty}, Yann and {Martino}, Nicolas and {Galli}, Richard and {Ludgate}, Mathew and {Blauensteiner}, Markus and {Reich}, Wolfgang and {Walker}, Sean and {di Cicco}, Dennis and {Mittelman}, David and {Morgan}, Curtis and {Kaeouach}, Aziz Ettahar and {Rupert}, Justin and {Benkhaldoun}, Zouhair},
        title = "{Deep Optical Emission-line Images of Nine Known and Three New Galactic Supernova Remnants}",
      journal = {\apjs},
         year = 2024,
        month = jun,
       volume = {272},
       number = {2},
          eid = {36},
        pages = {36},
          doi = {10.3847/1538-4365/ad410a},
archivePrefix = {arXiv},
       eprint = {2403.00317},
 primaryClass = {astro-ph.HE},
       adsurl = {https://ui.adsabs.harvard.edu/abs/2024ApJS..272...36F}
}

@ARTICLE{Kothes_2017A&A...597A.116K,
       author = {{Kothes}, Roland and {Reich}, Patricia and {Foster}, Tyler J. and {Reich}, Wolfgang},
        title = "{G181.1+9.5, a new high-latitude low-surface brightness supernova remnant}",
      journal = {\aap},
         year = 2017,
        month = jan,
       volume = {597},
          eid = {A116},
        pages = {A116},
          doi = {10.1051/0004-6361/201629848},
archivePrefix = {arXiv},
       eprint = {1612.01956},
 primaryClass = {astro-ph.HE},
       adsurl = {https://ui.adsabs.harvard.edu/abs/2017A&A...597A.116K}
}

@ARTICLE{Brogan_2006,
       author = {{Brogan}, C.~L. and {Gelfand}, J.~D. and {Gaensler}, B.~M. and {Kassim}, N.~E. and {Lazio}, T.~J.~W.},
        title = "{Discovery of 35 New Supernova Remnants in the Inner Galaxy}",
      journal = {\apjl},
         year = 2006,
        month = mar,
       volume = {639},
       number = {1},
        pages = {L25-L29},
          doi = {10.1086/501500},
archivePrefix = {arXiv},
       eprint = {astro-ph/0601451},
 primaryClass = {astro-ph},
       adsurl = {https://ui.adsabs.harvard.edu/abs/2006ApJ...639L..25B}
}

@ARTICLE{Hotan_2021,
       author = {{Hotan}, A.~W. and {Bunton}, J.~D. and {Chippendale}, A.~P. and {Whiting}, M. and {Tuthill}, J. and {Moss}, V.~A. and {McConnell}, D. and {Amy}, S.~W. and {Huynh}, M.~T. and {Allison}, J.~R. and {Anderson}, C.~S. and {Bannister}, K.~W. and {Bastholm}, E. and {Beresford}, R. and {Bock}, D.~C. -J. and {Bolton}, R. and {Chapman}, J.~M. and {Chow}, K. and {Collier}, J.~D. and {Cooray}, F.~R. and {Cornwell}, T.~J. and {Diamond}, P.~J. and {Edwards}, P.~G. and {Feain}, I.~J. and {Franzen}, T.~M.~O. and {George}, D. and {Gupta}, N. and {Hampson}, G.~A. and {Harvey-Smith}, L. and {Hayman}, D.~B. and {Heywood}, I. and {Jacka}, C. and {Jackson}, C.~A. and {Jackson}, S. and {Jeganathan}, K. and {Johnston}, S. and {Kesteven}, M. and {Kleiner}, D. and {Koribalski}, B.~S. and {Lee-Waddell}, K. and {Lenc}, E. and {Lensson}, E.~S. and {Mackay}, S. and {Mahony}, E.~K. and {McClure-Griffiths}, N.~M. and {McConigley}, R. and {Mirtschin}, P. and {Ng}, A.~K. and {Norris}, R.~P. and {Pearce}, S.~E. and {Phillips}, C. and {Pilawa}, M.~A. and {Raja}, W. and {Reynolds}, J.~E. and {Roberts}, P. and {Roxby}, D.~N. and {Sadler}, E.~M. and {Shields}, M. and {Schinckel}, A.~E.~T. and {Serra}, P. and {Shaw}, R.~D. and {Sweetnam}, T. and {Troup}, E.~R. and {Tzioumis}, A. and {Voronkov}, M.~A. and {Westmeier}, T.},
        title = "{Australian square kilometre array pathfinder: I. system description}",
      journal = {\pasa},
         year = 2021,
        month = mar,
       volume = {38},
          eid = {e009},
        pages = {e009},
          doi = {10.1017/pasa.2021.1},
archivePrefix = {arXiv},
       eprint = {2102.01870},
 primaryClass = {astro-ph.IM},
       adsurl = {https://ui.adsabs.harvard.edu/abs/2021PASA...38....9H}
}

@ARTICLE{Norris_2011,
       author = {{Norris}, Ray P. and {Hopkins}, A.~M. and {Afonso}, J. and {Brown}, S. and {Condon}, J.~J. and {Dunne}, L. and {Feain}, I. and {Hollow}, R. and {Jarvis}, M. and {Johnston-Hollitt}, M. and {Lenc}, E. and {Middelberg}, E. and {Padovani}, P. and {Prandoni}, I. and {Rudnick}, L. and {Seymour}, N. and {Umana}, G. and {Andernach}, H. and {Alexander}, D.~M. and {Appleton}, P.~N. and {Bacon}, D. and {Banfield}, J. and {Becker}, W. and {Brown}, M.~J.~I. and {Ciliegi}, P. and {Jackson}, C. and {Eales}, S. and {Edge}, A.~C. and {Gaensler}, B.~M. and {Giovannini}, G. and {Hales}, C.~A. and {Hancock}, P. and {Huynh}, M.~T. and {Ibar}, E. and {Ivison}, R.~J. and {Kennicutt}, R. and {Kimball}, Amy E. and {Koekemoer}, A.~M. and {Koribalski}, B.~S. and {L{\'o}pez-S{\'a}nchez}, {\'A}. R. and {Mao}, M.~Y. and {Murphy}, T. and {Messias}, H. and {Pimbblet}, K.~A. and {Raccanelli}, A. and {Randall}, K.~E. and {Reiprich}, T.~H. and {Roseboom}, I.~G. and {R{\"o}ttgering}, H. and {Saikia}, D.~J. and {Sharp}, R.~G. and {Slee}, O.~B. and {Smail}, Ian and {Thompson}, M.~A. and {Urquhart}, J.~S. and {Wall}, J.~V. and {Zhao}, G. -B.},
        title = "{EMU: Evolutionary Map of the Universe}",
      journal = {\pasa},
         year = 2011,
        month = aug,
       volume = {28},
       number = {3},
        pages = {215-248},
          doi = {10.1071/AS11021},
archivePrefix = {arXiv},
       eprint = {1106.3219},
 primaryClass = {astro-ph.CO},
       adsurl = {https://ui.adsabs.harvard.edu/abs/2011PASA...28..215N}
}

@ARTICLE{Norris_2021,
       author = {{Norris}, Ray P. and {Marvil}, Joshua and {Collier}, J.~D. and {Kapi{\'n}ska}, Anna D. and {O'Brien}, Andrew N. and {Rudnick}, L. and {Andernach}, Heinz and {Asorey}, Jacobo and {Brown}, Michael J.~I. and {Br{\"u}ggen}, Marcus and {Crawford}, Evan and {English}, Jayanne and {Rahman}, Syed Faisal ur and {Filipovi{\'c}}, Miroslav D. and {Gordon}, Yjan and {G{\"u}rkan}, G{\"u}lay and {Hale}, Catherine and {Hopkins}, Andrew M. and {Huynh}, Minh T. and {HyeongHan}, Kim and {James Jee}, M. and {Koribalski}, B{\"a}rbel S. and {Lenc}, Emil and {Luken}, Kieran and {Parkinson}, David and {Prandoni}, Isabella and {Raja}, Wasim and {Reiprich}, Thomas H. and {Riseley}, Christopher J. and {Shabala}, Stanislav S. and {Sheil}, Jaimie R. and {Vernstrom}, Tessa and {Whiting}, Matthew T. and {Allison}, James R. and {Anderson}, C.~S. and {Ball}, Lewis and {Bell}, Martin and {Bunton}, John and {Galvin}, T.~J. and {Gupta}, Neeraj and {Hotan}, Aidan and {Jacka}, Colin and {Macgregor}, Peter J. and {Mahony}, Elizabeth K. and {Maio}, Umberto and {Moss}, Vanessa and {Pandey-Pommier}, M. and {Voronkov}, Maxim A.},
        title = "{The Evolutionary Map of the Universe pilot survey}",
      journal = {\pasa},
         year = 2021,
        month = sep,
       volume = {38},
          eid = {e046},
        pages = {e046},
          doi = {10.1017/pasa.2021.42},
archivePrefix = {arXiv},
       eprint = {2108.00569},
 primaryClass = {astro-ph.CO},
       adsurl = {https://ui.adsabs.harvard.edu/abs/2021PASA...38...46N}
}

@ARTICLE{Gaensler_2025PASA...42...91G,
       author = {{Gaensler}, B.~M. and {Heald}, G.~H. and {McClure-Griffiths}, N.~M. and {Anderson}, C.~S. and {Van Eck}, C.~L. and {West}, J.~L. and {Thomson}, A.~J.~M. and {Leahy}, J.~P. and {Rudnick}, L. and {Ma}, Y.~K. and {Akahori}, Takuya and {G{\"u}rkan}, G. and {Landecker}, T.~L. and {Mao}, S.~A. and {O'Sullivan}, S.~P. and {Raja}, W. and {Sun}, X. and {Vernstrom}, T. and {Baidoo}, Lerato and {Carretti}, Ettore and {Taylor}, A.~R. and {Willis}, A.~G. and {Osinga}, Erik and {Livingston}, J.~D. and {Alexander}, E.~L. and {Alonso-L{\'o}pez}, David and {Amaral}, A.~D. and {An}, T. and {Bracco}, Andrea and {Bradbury}, S. and {Br{\"u}ggen}, Marcus and {Eswaraiah}, Chakali and {En{\ss}lin}, Torsten and {Galvin}, T.~J. and {Haverkorn}, Marijke and {Hopkins}, A.~M. and {Hutschenreuter}, Sebastian and {Ideguchi}, Shinsuke and {Jaswanth}, S. and {Jung}, S. Lyla and {Kaczmarek}, J.~F. and {Kothes}, Roland and {Lazarevi{\'c}}, Sanja and {Leahy}, Denis and {Loi}, Francesca and {Marvil}, Joshua R. and {Norris}, Ray and {Pandhi}, Ayush and {Price}, Jason M. and {Riseley}, C.~J. and {Ryder}, P. and {Seta}, Amit and {Shaw}, Vasundhara and {Shen}, A.~X. and {Sobey}, C. and {Stil}, J. and {Stuardi}, Chiara and {Upasana}, Gupta and {Vanderwoude}, Shannon and {Velovi{\'c}}, Velibor},
        title = "{The Polarisation Sky Survey of the Universe's Magnetism (POSSUM): Science goals and survey description}",
      journal = {\pasa},
         year = 2025,
        month = jun,
       volume = {42},
          eid = {e091},
        pages = {e091},
          doi = {10.1017/pasa.2025.10031},
archivePrefix = {arXiv},
       eprint = {2505.08272},
 primaryClass = {astro-ph.GA},
       adsurl = {https://ui.adsabs.harvard.edu/abs/2025PASA...42...91G}
}

@INPROCEEDINGS{Gaensler_2010,
       author = {{Gaensler}, Bryan M. and {Landecker}, T.~L. and {Taylor}, A.~R. and {POSSUM Collaboration}},
        title = "{Survey Science with ASKAP: Polarization Sky Survey of the Universe's Magnetism (POSSUM)}",
    booktitle = {American Astronomical Society Meeting Abstracts \#215},
         year = 2010,
       series = {American Astronomical Society Meeting Abstracts},
       volume = {215},
        month = jan,
          eid = {470.13},
        pages = {470.13},
       adsurl = {https://ui.adsabs.harvard.edu/abs/2010AAS...21547013G}
}

@ARTICLE{Hopkins_2025PASA...42...71H,
       author = {{Hopkins}, Andrew and {Kapinska}, Anna and {Marvil}, Joshua and {Vernstrom}, Tessa and {Collier}, Jordan and {Norris}, Ray and {Gordon}, Yjan and {Duchesne}, Stefan and {Rudnick}, Lawrence and {Gupta}, Nikhel and {Carretti}, Ettore and {Anderson}, Craig and {Dai}, Shi and {G{\"u}rkan}, Gulay and {Parkinson}, David and {Prandoni}, Isabella and {Riggi}, Simone and {Shekhar Saraf}, Chandra and {Ma}, Yik Ki and {Filipovi{\'c}}, Miroslav D. and {Umana}, Grazia and {Bahr-Kalus}, Benedict and {Koribalski}, B{\"a}rbel Silvia and {Lenc}, Emil and {Ingallinera}, Adriano and {Afonso}, Jos{\'e} and {Ahmad}, Adeel and {Ahmed}, Ummee Tania and {Alexander}, Emma and {Andernach}, Heinz and {Asorey}, Jacobo and {Battisti}, Andrew J. and {Bilicki}, Maciej and {Botteon}, Andrea and {Brown}, Michael and {Br{\"u}ggen}, Marcus and {Cowley}, Michael and {Dage}, Kristen and {Hale}, Catherine Laura and {Hardcastle}, Martin and {Kothes}, Roland and {Lazarevi{\'c}}, Sanja and {Lin}, Yen-Ting and {Luken}, Kieran and {Moss}, Jeremy and {Prathap}, P.~K. Jahang and {ur Rahman}, Syed Faisal and {Reiprich}, Thomas and {Riseley}, Christopher and {Salvato}, Mara and {Seymour}, Nicholas and {Shabala}, Stanislav and {Smith}, Daniel and {Vaccari}, Mattia and {van Loon}, Jacco Th. and {Wong}, O. Ivy Ivy and {Zainal Alsaberi}, Rami and {Asher}, Albany and {Ball}, Brianna and {Barbosa}, Davi and {Biava}, Nadia and {Bradley}, Aaron and {Carvajal}, Rodrigo and {Crawford}, Evan J. and {Galvin}, Timothy James and {Huynh}, Minh and {Leahy}, Denis and {Matute}, Israel and {Moss}, Vanessa and {Pappalardo}, Ciro and {Smeaton}, Zachary and {Velovi{\'c}}, Velibor and {Zafar}, Tayyaba},
        title = "{The Evolutionary Map of the Universe: A new radio atlas for the southern hemisphere sky}",
      journal = {\pasa},
         year = 2025,
        month = may,
       volume = {42},
          eid = {e071},
        pages = {e071},
          doi = {10.1017/pasa.2025.10042},
archivePrefix = {arXiv},
       eprint = {2505.08271},
 primaryClass = {astro-ph.GA},
       adsurl = {https://ui.adsabs.harvard.edu/abs/2025PASA...42...71H}
}

@misc{Guzman_2019,
       author = {{Guzman}, Juan and {Whiting}, Matthew and {Voronkov}, Max and {Mitchell}, Daniel and {Ord}, Stephen and {Collins}, Daniel and {Marquarding}, Malte and {Lahur}, Paulus and {Maher}, Tony and {Van Diepen}, Ger and {Bannister}, Keith and {Wu}, Xinyu and {Lenc}, Emil and {Khoo}, Jonathan and {Bastholm}, Eric},
        title = "{ASKAPsoft: ASKAP science data processor software}",
 howpublished = {Astrophysics Source Code Library, record ascl:1912.003},
         year = 2019,
        month = dec,
          eid = {ascl:1912.003},
       adsurl = {https://ui.adsabs.harvard.edu/abs/2019ascl.soft12003G}
}

@ARTICLE{Jing_2025ApJ...980..162J,
       author = {{Jing}, Wenhui and {West}, Jennifer L. and {Sun}, Xiaohui and {Raja}, Wasim and {Li}, Xianghua and {Dang}, Lingxiao and {Zhou}, Ping and {Filipovi{\'c}}, Miroslav D. and {Hopkins}, Andrew M. and {Kothes}, Roland and {Lazarevi{\'c}}, Sanja and {Leahy}, Denis and {Lenc}, Emil and {Ma}, Yik Ki and {Van Eck}, Cameron L.},
        title = "{ASKAP Observations of the Radio Shell in the Composite Supernova Remnant G310.6-1.6}",
      journal = {\apj},
         year = 2025,
        month = feb,
       volume = {980},
       number = {2},
          eid = {162},
        pages = {162},
          doi = {10.3847/1538-4357/adad66},
archivePrefix = {arXiv},
       eprint = {2503.01364},
 primaryClass = {astro-ph.GA},
       adsurl = {https://ui.adsabs.harvard.edu/abs/2025ApJ...980..162J}
}

@ARTICLE{Wright_2010,
       author = {{Wright}, Edward L. and {Eisenhardt}, Peter R.~M. and {Mainzer}, Amy K. and {Ressler}, Michael E. and {Cutri}, Roc M. and {Jarrett}, Thomas and {Kirkpatrick}, J. Davy and {Padgett}, Deborah and {McMillan}, Robert S. and {Skrutskie}, Michael and {Stanford}, S.~A. and {Cohen}, Martin and {Walker}, Russell G. and {Mather}, John C. and {Leisawitz}, David and {Gautier}, III, Thomas N. and {McLean}, Ian and {Benford}, Dominic and {Lonsdale}, Carol J. and {Blain}, Andrew and {Mendez}, Bryan and {Irace}, William R. and {Duval}, Valerie and {Liu}, Fengchuan and {Royer}, Don and {Heinrichsen}, Ingolf and {Howard}, Joan and {Shannon}, Mark and {Kendall}, Martha and {Walsh}, Amy L. and {Larsen}, Mark and {Cardon}, Joel G. and {Schick}, Scott and {Schwalm}, Mark and {Abid}, Mohamed and {Fabinsky}, Beth and {Naes}, Larry and {Tsai}, Chao-Wei},
        title = "{The Wide-field Infrared Survey Explorer (WISE): Mission Description and Initial On-orbit Performance}",
      journal = {\aj},
         year = 2010,
        month = dec,
       volume = {140},
       number = {6},
        pages = {1868-1881},
          doi = {10.1088/0004-6256/140/6/1868},
archivePrefix = {arXiv},
       eprint = {1008.0031},
 primaryClass = {astro-ph.IM},
       adsurl = {https://ui.adsabs.harvard.edu/abs/2010AJ....140.1868W}
}

@Article{Hunter_2007,
  Author    = {Hunter, J. D.},
  Title     = {Matplotlib: A 2D graphics environment},
  Journal   = {Computing in Science \& Engineering},
  Volume    = {9},
  Number    = {3},
  Pages     = {90--95},
  publisher = {IEEE COMPUTER SOC},
  doi       = {10.1109/MCSE.2007.55},
  year      = 2007
}

@article{Bell_1978,
    author = {Bell, A. R.},
    title = "{The acceleration of cosmic rays in shock fronts – I}",
    journal = {\mnras},
    volume = {182},
    number = {2},
    pages = {147-156},
    year = {1978},
    month = {02},
    issn = {0035-8711},
    doi = {10.1093/mnras/182.2.147},
    url = {https://doi.org/10.1093/mnras/182.2.147},
    eprint = {https://academic.oup.com/mnras/article-pdf/182/2/147/3710138/mnras182-0147.pdf},
}

@ARTICLE{Blandford_1978,
       author = {{Blandford}, R.~D. and {Ostriker}, J.~P.},
        title = "{Particle acceleration by astrophysical shocks.}",
      journal = {\apjl},
         year = 1978,
        month = apr,
       volume = {221},
        pages = {L29-L32},
          doi = {10.1086/182658},
       adsurl = {https://ui.adsabs.harvard.edu/abs/1978ApJ...221L..29B}
}

@ARTICLE{Whiteoak_1996,
       author = {{Whiteoak}, J.~B.~Z. and {Green}, A.~J.},
        title = "{The MOST supernova remnant catalogue (MSC).}",
      journal = {\aaps},
         year = 1996,
        month = aug,
       volume = {118},
        pages = {329-380},
       adsurl = {https://ui.adsabs.harvard.edu/abs/1996A&AS..118..329W}
}

@ARTICLE{Voges_1999,
       author = {{Voges}, W. and {Aschenbach}, B. and {Boller}, Th. and {Br{\"a}uninger}, H. and {Briel}, U. and {Burkert}, W. and {Dennerl}, K. and {Englhauser}, J. and {Gruber}, R. and {Haberl}, F. and {Hartner}, G. and {Hasinger}, G. and {K{\"u}rster}, M. and {Pfeffermann}, E. and {Pietsch}, W. and {Predehl}, P. and {Rosso}, C. and {Schmitt}, J.~H.~M.~M. and {Tr{\"u}mper}, J. and {Zimmermann}, H.~U.},
        title = "{The ROSAT all-sky survey bright source catalogue}",
      journal = {\aap},
         year = 1999,
        month = sep,
       volume = {349},
        pages = {389-405},
          doi = {10.48550/arXiv.astro-ph/9909315},
archivePrefix = {arXiv},
       eprint = {astro-ph/9909315},
 primaryClass = {astro-ph},
       adsurl = {https://ui.adsabs.harvard.edu/abs/1999A&A...349..389V}
}

@ARTICLE{Berezhko_2004,
       author = {{Berezhko}, E.~G. and {V{\"o}lk}, H.~J.},
        title = "{The theory of synchrotron emission from supernova remnants}",
      journal = {\aap},
         year = 2004,
        month = nov,
       volume = {427},
        pages = {525-536},
          doi = {10.1051/0004-6361:20041111},
archivePrefix = {arXiv},
       eprint = {astro-ph/0408121},
 primaryClass = {astro-ph},
       adsurl = {https://ui.adsabs.harvard.edu/abs/2004A&A...427..525B}
}

@ARTICLE{Green_2005,
       author = {{Green}, D.~A.},
        title = "{Some statistics of Galactic SNRs}",
      journal = {\memsai},
         year = 2005,
        month = jan,
       volume = {76},
        pages = {534-541},
          doi = {10.48550/arXiv.astro-ph/0505428},
archivePrefix = {arXiv},
       eprint = {astro-ph/0505428},
 primaryClass = {astro-ph},
       adsurl = {https://ui.adsabs.harvard.edu/abs/2005MmSAI..76..534G}
}

@ARTICLE{Shklovskii_1960,
       author = {{Shklovskii}, I.~S.},
        title = "{Secular Variation of the Flux and Intensity of Radio Emission from Discrete Sources}",
      journal = {\sovast},
         year = 1960,
        month = oct,
       volume = {4},
        pages = {243},
       adsurl = {https://ui.adsabs.harvard.edu/abs/1960SvA.....4..243S}
}

@ARTICLE{Pavlovic_2014,
       author = {{Pavlovic}, M.~Z. and {Dobardzic}, A. and {Vukotic}, B. and {Urosevic}, D.},
        title = "{Updated Radio Sigma-D Relation for Galactic Supernova Remnants}",
      journal = {Serb.~Astron.~J.},
         year = 2014,
        month = dec,
       volume = {189},
        pages = {25-40},
          doi = {10.2298/SAJ1489025P},
archivePrefix = {arXiv},
       eprint = {1411.2234},
 primaryClass = {astro-ph.HE},
       adsurl = {https://ui.adsabs.harvard.edu/abs/2014SerAJ.189...25P}
}

@ARTICLE{Lupton_2004,
       author = {{Lupton}, Robert and {Blanton}, Michael R. and {Fekete}, George and {Hogg}, David W. and {O'Mullane}, Wil and {Szalay}, Alex and {Wherry}, Nicholas},
        title = "{Preparing Red-Green-Blue Images from CCD Data}",
      journal = {\pasp},
         year = 2004,
        month = feb,
       volume = {116},
       number = {816},
        pages = {133-137},
          doi = {10.1086/382245},
archivePrefix = {arXiv},
       eprint = {astro-ph/0312483},
 primaryClass = {astro-ph},
       adsurl = {https://ui.adsabs.harvard.edu/abs/2004PASP..116..133L}
}

@misc{Lazarevic_in_prep,
    author  = "Sanja Lazarevi{\'c}",
    year    = "\noop{3001} et al., in prep.",
}

@BOOK{Longair_2011hea..book.....L,
       author = {{Longair}, Malcolm S.},
        title = "{High Energy Astrophysics}",
         year = 2011,
       adsurl = {https://ui.adsabs.harvard.edu/abs/2011hea..book.....L}
}
%


%
%
\end{document}